\documentstyle[11pt,epsfig]{article}
\input{epsf}
\begin{document}
\setlength{\baselineskip}{0.18in}
\newcommand{\be}{\begin{eqnarray}}
\newcommand{\ee}{\end{eqnarray}}
\newcommand{\bi}{\bibitem}
\newcommand{\lar}{\leftarrow}
\newcommand{\rar}{\rightarrow}
\newcommand{\lrar}{\leftrightarrow}
\newcommand{\mpl}{m_{Pl}^2}
\newcommand{\mplq}{m_{Pl}}
\newcommand{\rmn}{R_{\mu\nu}}
\newcommand{\gmn}{g_{\mu\nu}}
\newcommand{\non}{\nonumber}  
\newcommand{\del}{\partial} 
\newcommand{\la}{\left\langle} 
\newcommand{\ra}{\right\rangle}

\newcommand{\vmm}{V_{\mu\mu}}
\newcommand{\vtt}{V_{\tau\tau}}
\newcommand{\vmt}{V_{\mu\tau}}
\newcommand{\nue}{\nu_e}
\newcommand{\num}{\nu_\mu}
\newcommand{\nut}{\nu_\tau}
\newcommand{\nus}{\nu_s}
\newcommand{\nua}{\nu_a}
\newcommand{\mne}{m_{\nu_e}}
\newcommand{\mnm}{m_{\nu_\mu}}
\newcommand{\mnt}{m_{\nu_\tau}}
\newcommand{\nuh}{\nu_h}
\newcommand{\mnh}{m_{\nu_h}}
\newcommand{\taut}{\tau_{\nut}}
\newcommand{\fg}{f_{\gamma}}
\newcommand{\dE}{{\delta E}}
\newcommand{\dm}{{\delta m^2}}
\newcommand{\rmt}{\rho_{\mu\tau}}
\newcommand{\rtm}{\rho_{\tau\mu}}
\newcommand{\rba}{\rho_{ba}}
\newcommand{\rab}{\rho_{ab}}
\newcommand{\raa}{\rho_{aa}}
\newcommand{\rcb}{\rho_{cb}}
\newcommand{\rac}{\rho_{ac}}
\newcommand{\rtt}{\rho_{\tau\tau}}
\newcommand{\rbs}{\rho_{bs}}
\newcommand{\htt}{{\cal H}_{\tau\tau}}
\newcommand{\htm}{{\cal H}_{\tau\mu}}
\newcommand{\hee}{{\cal H}_{ee}}
\newcommand{\hmm}{{\cal H}_{\mu\mu}}
\newcommand{\hem}{{\cal H}_{e\mu}}
\newcommand{\hes}{{\cal H}_{es}}
\newcommand{\hms}{{\cal H}_{\mu s}}
\newcommand{\hmt}{{\cal H}_{\mu \tau}}
\newcommand{\het}{{\cal H}_{e \tau}}
\newcommand{\hts}{{\cal H}_{\tau s}}
\newcommand{\hab}{{\cal H}_{ab}}

\newcommand{\dms}{\Delta_{\mu s}}
\newcommand{\dme}{\Delta_{\mu e}}
\newcommand{\dmt}{\Delta_{\mu \tau}}
\newcommand{\des}{\Delta_{e s}}

\newcommand{\deta}{\Delta_{e \tau}}
\newcommand{\dts}{\Delta_{\tau s}}
\newcommand{\dsm}{\Delta_{s\mu }}
\newcommand{\dem}{\Delta_{e\mu}}
\newcommand{\dtm}{\Delta_{\tau \mu}}
\newcommand{\dse}{\Delta_{se}}
\newcommand{\dte}{\Delta_{\tau e}}
\newcommand{\dst}{\Delta_{s\tau }}

\newcommand{\gee}{\gamma_{ee}}
\newcommand{\gem}{\gamma_{\mu\mu}}
\newcommand{\ges}{\gamma_{es}}
\newcommand{\gms}{\gamma_{\mu s}}
\newcommand{\rem}{\rho_{e\mu}}
\newcommand{\rme}{\rho_{\mu e}}
\newcommand{\res}{\rho_{es}}
\newcommand{\rse}{\rho_{se}}
\newcommand{\rms}{\rho_{\mu s}}
\newcommand{\rsm}{\rho_{s\mu}}
\newcommand{\rmm}{\rho_{\mu \mu}}
\newcommand{\ree}{\rho_{ee}}

\newcommand{\gtrsim}{ \mathop{}_{\textstyle \sim}^{\textstyle >} }
\newcommand{\lesssim}{ \mathop{}_{\textstyle \sim}^{\textstyle <} }
\newcommand{\ds}{\displaystyle}

\begin{center}
\vglue .06in
{\Large \bf { 
Do neutrino flavor oscillations forbid large lepton asymmetry of
the universe?
  }
}
\bigskip
\\{\bf A.D. Dolgov}$^{(a)(b)(c)}$ and 
{\bf Fuminobu Takahashi}$^{(d)}$
\\[.2in]
$^{(a)}${\it INFN, sezione di Ferrara,
Via Paradiso, 12 - 44100 Ferrara,
Italy} \\
$^{(b)}${\it ITEP, Bol. Cheremushkinskaya 25, Moscow 113259, Russia.
}  \\
$^{(c)}${\it ICTP, Trieste, 34014, Italy
}  \\
$^{(d)}${\it Research Center for the Early Universe, Graduate School of 
Science, \\University of Tokyo, Tokyo 113-0033, Japan
}

\end{center}

\vspace{.3in}
\begin{abstract}
It is shown that hypothetical neutrino-majoron coupling can suppress
neutrino flavor oscillations in the early universe, in contrast to 
the usual weak interaction case. This reopens a window for a noticeable 
cosmological lepton asymmetry which is forbidden for the large mixing 
angle solution in the case of standard interactions of neutrinos.  
\end{abstract}

\bigskip

\section{Introduction \label{s-intr}}

Cosmological lepton asymmetry is not directly measurable, in contrast
to baryon asymmetry, but may be observed or restricted through its
impact on big bang nucleosynthesis (BBN), large scale structure
formation, and the angular spectrum of the cosmic microwave background
radiation (CMBR), for a review see e.g. Ref.~\cite{dolgov02}.  At the
present time the best bounds follow from the consideration of
BBN. Primordial production of light elements is especially sensitive
to the value of asymmetry between electronic neutrinos and
antineutrinos since they directly influence the neutron-proton
transformations in weak reactions $\nue\,n \lrar e^- p$ and $\bar\nue
\,p \lrar e^+ n$. The bound on the chemical potential of electronic
neutrinos obtained in Ref.~\cite{kohri96} reads $|\xi_e| \equiv
|\mu_e/T| < 0.1$. The bound on muonic or tauonic asymmetry is
noticeably weaker because $\num$ or $\nut$ produce an effect on BBN
only through their impact on the cooling rate of the universe and the
degeneracy of these neutrino flavors is equivalent to an addition of
\be
\Delta N_\nu = \frac{15}{7}\,\left[\left(\frac{\xi_{\mu,\tau}}{\pi}
\right)^4 + 2\,\left(\frac{\xi_{\mu,\tau}}{\pi}\right)^2\right]
\label{delta-nu}
\ee
massless neutrino species at the BBN epoch. If the latter is bounded
by $\Delta N_\nu <1$ (safe bound) then $|\xi_{\mu,\tau}|< 1.5$. For
less conservative bound, $\Delta N_\nu <0.4$~\cite{Barger:2003zg}, the
dimensionless chemical potentials should be below 0.9.  Still,
asymmetry of the order of unity would have noticeable cosmological
effects on large scale structure formation and CMBR.  The limits
quoted above are valid if the chemical potential of only one kind of
neutrino is non-zero. If a conspiracy between different chemical
potentials is allowed, such that a positive effect of $\xi_\mu$ or
$\xi_\tau$ is compensated by $\xi_e$ or vice versa, the limits would
be somewhat weaker. According to Ref.~\cite{hansen01} they are:
$|\xi_e|<0.2$ and $|\xi_{\mu,\tau}|<2.6$. \footnote{We are not
concerned in this paper with the origin of such large lepton
asymmetry. Thus far, many cosmological scenarios that explain both the
small observed baryon asymmetry and a possible large lepton asymmetry
simultaneously, have been proposed~\cite{largeL_smallB}.}

The bounds on chemical potentials of $\num$ and $\nut$ can be
significantly improved because of the strong mixing between different
neutrino flavors~\cite{nu-mixing}. This mixing gives rise to the fast
transformation between $\nue$, $\num$, and $\nut$ in the early
universe and leads to equilibration of asymmetries of all neutrino
species. Thus the BBN bound on any chemical potential becomes
essentially that obtained for $\nue$~\cite{act-act} (see also the
papers~\cite{lunardini01}):
\be
|\xi_{e,\mu,\tau}| < 0.07.
\label{xiemt}
\ee
In this case the cosmological impact of neutrino degeneracy would be
negligible. However, a compensation of $\xi_e$ by other types of
radiation is still possible, which was studied in
Ref.~\cite{Barger:2003rt}.  The authors of Ref.~\cite{Barger:2003rt}
put the constraints on $\xi_e$ and $\Delta N_\nu$ based on WMAP data
and BBN:
\be
-0.1 \lesssim \xi_{e} \lesssim 0.3,~~-2 \lesssim \Delta N_\nu \lesssim 5.
\label{extra_rad}
\ee

It is interesting to see if one could reasonably modify the standard
model to allow large muonic and/or tauonic charge asymmetries,
together with a small electronic asymmetry, to avoid conflict with
BBN. This is the aim of this work. A natural generalization is to
introduce an additional interaction of neutrinos with massless or
light (pseudo)Nambu-Goldstone boson, majoron~\cite{majoron}. The idea
to invoke majoron to modify neutrino oscillations in primeval plasma
was discussed in Refs.~\cite{babu92,bento01}. In these papers, two
different mechanisms which could block or suppress oscillations
between active and hypothetical sterile neutrinos have been
considered.  We have, however, some concerns about the validity of
their results which we will discuss in the next section. Let us note
that in this paper we consider an impact of neutrino majoron
interactions on the oscillations between active neutrinos and not on
active-sterile oscillations, as is done in the mentioned above papers.

\section{General discussion and approximate estimates \label{s-gen}}

Neutrino oscillations may be suppressed in medium if the interaction
of neutrinos with the medium is sufficiently strong or in other words
the refraction index, $n$ (or effective potential, $V=E(n-1)$) of
neutrinos is very large. Correspondingly the mixing angle in medium
$\theta_m$ becomes negligible:
\be
\tan^2 2\theta_m = 
{s_2^2 \over \left( c_2^2  + V /\delta E \right)^2 +  
\Gamma^2 /(4\delta E^2) } 
\label{sintm}
\ee
where $s_2=\sin 2\theta$, $c_2 =\cos 2\theta$, $\theta$ is the vacuum
mixing angle, $\delta E = \dm /2E$, $E$ is the neutrino energy, and
$\Gamma$ is the rate of neutrino interaction with medium given by
\be
\Gamma = \frac {80 (1+g_L^2+g_R^2)\,G^2_F E T^4}{3\pi^3}
\label{gamma}
\ee
where $T$ is the temperature of the primeval plasma, $G_F= 1.166\cdot
10^{-11}\,{\rm MeV}^{-2}$ is the Fermi coupling constant, $g_L =
\sin^2 \theta_W \pm 1/2$ and $g_R = \sin^2 \theta_W$ where $\sin^2
\theta_W =0.23$ and the sign $''+''$ stands for $\nue$ and $''-''$
stands for $\nu_{\mu,\tau}$.  The derivation of these equations can be
found e.g. in lectures~\cite{ad-itep02}.

The diagonal components of the effective potential, created by the
standard weak interaction, for the active neutrino species are given
by~\cite{nora}:
\be
V_{aa}^{(w)} = \pm C_1 \eta^{(a)} G_FT^3 -
C_2^a \frac{G^2_F T^4 E}{\alpha}
\label{veff}
\ee
where $a=e,\mu,\tau$ labels the neutrino flavors, $\alpha=1/137$ is
the fine structure constant, and the signs ``$\pm$'' refer to
neutrinos and antineutrinos respectively.  The first term arises due
to a possible charge asymmetry of the primeval plasma, while the
second one comes from the non-locality of weak interactions associated
with the exchange of $W$ or $Z$ bosons.  According to Ref.~\cite{nora}
the coefficients $C_j$ are: $C_1 \approx 0.95$, $C_2^e \approx 0.62$
and $C_2^{\mu,\tau} \approx 0.17$ (for $T<m_\mu$).  These values are
true in the limit of thermal equilibrium, otherwise these coefficients
are some integrals from the distribution functions over momenta.  The
charge asymmetry of plasma is described by the coefficients
$\eta^{(a)}$ which are equal to
\be
\eta^{(e)}& =&
2\eta_{\nue} +\eta_{\num} + \eta_{\nut} +\eta_{e}-\eta_{n}/2 \,\,\,
 ( {\rm for} \,\, \nue)~,
\label{etanue} \\
\eta^{(\mu)} &=&
2\eta_{\num} +\eta_{\nue} + \eta_{\nut} - \eta_{n}/2\,\,\,
({\rm for} \,\, \num)~,
\label{etanumu}
\ee
and $\eta^{(\tau)}$ for $\nut$ is obtained from Eq.~(\ref{etanumu}) by
the interchange $\mu \lrar \tau$. The individual charge asymmetries,
$\eta_X$, are defined as the ratio of the difference between
particle-antiparticle number densities to the number density of
photons with the account of the 11/4-factor emerging from the
$e^+e^-$-annihilation:
\be
\eta_X = \left(N_X -N_{\bar X}\right) /N_\gamma.
\label{etax}
\ee

For ``normal'' values of charge asymmetry, i.e. $|\eta| \sim 10^{-9}$
the charge asymmetric term in the potential (\ref{veff}) is
subdominant but if the asymmetry is of the order of unity, then the
ratio $V_{aa}/\dE$ becomes huge:
\be
V_{aa}^{(w)}/\dE \approx 2\cdot 10^{-11} \eta E T^3 (\dm)^{-1} 
 \left({\rm MeV}\right)^{-2}
\approx 20 \eta \left(T/{\rm MeV}\right)^4 ({\rm eV}^2/\dm)
\label{vdE}
\ee
and e.g. for $\dm = 10^{-2}$ eV$^2$ and $\eta \sim 1$ the mixing angle
in matter would be suppressed more than by three orders of magnitude
in the BBN range of temperatures, $T\sim$MeV.

However, this result is valid for mixing between active and sterile
neutrinos and is not true for active-active mixing. In the last case
the effective potential has large off-diagonal matrix
elements~\cite{v-off}, $V_{ab}$, with $a\neq b$, which compensate the
suppression induced by the diagonal terms. This is the reason for
non-suppressed oscillations between active flavors and for the
equilibration of all leptonic asymmetries~\cite{act-act}.

Possible additional interactions of neutrinos with majoron can be
described by the Lagrangian:
\be 
{\cal L}_{int} \sim \chi (g_{ab}\nu_a^T C \nu_b + h.c )
\label{lint}
\ee
where $\chi$ is the operator of the majoron field, $C$ is the matrix of
charge conjugation, and $g_{ab}$ are the coupling constants. We will 
assume for simplicity, though it is not necessary, that only 
the flavor-diagonal coupling is non-vanishing, 
$g_{ab}\sim g_{aa}\delta_{ab}$. We will consider the coupling matrix
with more general form later. 
Let us also assume that the majoron is
so light that its decay and inverse decay are not essential at the
BBN epoch. Neutrino-(anti)neutrino scattering through majoron exchange
gives rise to a contribution to neutrino effective potential 
(proportional to forward scattering amplitude) which can be estimated 
as
\be
V^{(\chi)}_{ab} \sim g_{aa} g_{bb} T.
\label{vjab}
\ee

The constants $g_{aa}$ should satisfy several constraints to make the
mechanism operative. Firstly, the diagonal part of the potential
$V^{(\chi)}_{aa}$ should be larger than the weak potential $V^{(w)}$
given by Eq.~(\ref{veff}), while its off-diagonal components must be
much smaller than the diagonal ones,
$V^{(\chi)}_{ab}<<V^{(\chi)}_{aa}$, that is, the flavor symmetry in
the neutrino-majoron interactions should be strongly broken. These two
conditions would ensure suppression of flavor changing oscillations
between active neutrinos.  To realize these conditions the following
constraints should be imposed:
\be
g_{aa}^2 > 10^{-11}\eta^{(a)}\, \left(\frac{T_{ae}}{1{\rm~ MeV}}\right)^2
\,\,\, {\rm and}\,\,\,
g_{ee} \ll g_{aa}.
\label{gaa-gee}
\ee
Here $a$ labels $\mu$ or $\tau$ and the coupling of majoron to $\nue$
is assumed to be much weaker than those to $\num$ and/or $\nut$;
$T_{ae}$ is the temperature at which $\nu_a$ are effectively
transformed into $\nue$ in the standard theory. According to
calculations of Ref.~\cite{act-act} it takes place around $T_{ae} = 1$
MeV for the large mixing angle solution to solar neutrino deficit,
while the transformation between $\num$ and $\nut$ in the standard
theory takes place somewhat above 10 MeV. In fact, as we see in what
follows from numerical calculations, the more accurate bound is much
less restrictive than this simple estimate (see Eq.~(\ref{dL/dx}) or
Figs.~\ref{fig:diagram6} and \ref{fig:diagram7} below).

The second inequality (\ref{gaa-gee}) implies that the off-diagonal
components of the effective potential are suppressed in comparison with
the diagonal ones and the oscillations remain blocked. This is not
so in the case of the standard weak interactions which are flavor
symmetric and the large value of the denominator due to large 
$V_{aa}$ in Eq.~(\ref{sintm}) is compensated by the same large off-diagonal 
components of the effective potential. 

Secondly, the coupling constants $g_{aa}$ should not be too large,
otherwise flavor non-conserving reactions of the type 
$\nue\,\nu_a \lrar \bar \nue \bar\nu_a$ (or similar) would lead
to equilibration of all leptonic charges. To avoid that the rate 
of these reactions, $\Gamma_{ea} \sim \sigma_{ea} T^3$, should be
smaller than the cosmological expansion rate $H\sim T^2/m_{Pl}$, 
where $m_{Pl} = 1.221\cdot 10^{22}$ MeV is the Planck mass. Thus,
to suppress $e-\mu$ or $e-\tau$ transformation through direct 
reactions one needs
\be
g_{aa}^2g_{ee}^2 <10^{-22} \left(\frac{T}{1{\rm~ MeV}}\right).
\label{g-react}
\ee
This conditions should be satisfied for temperatures above the BBN
range, i.e. $T>1$ MeV. Similarly, if we require that $\nu_a\,\nu_a
\lrar \bar \nu_a \bar\nu_a$ should not occur efficiently, the coupling
constants must satisfy a similar inequality with $g_{ee}$ replaced
with $g_{aa}$.

There are quite strong limits on possible coupling of majoron to
neutrinos which follow from astrophysics~\cite{tomas01,farzan02};
discussion and references to earlier works can be found e.g. in the
book~\cite{raffelt}. Astrophysics allows either very small or quite
large coupling constants. The former is quite evident, while the
latter appears because strongly interacting majorons, though
efficiently produced inside a star, cannot propagate out and carry
away the energy, thus opening a window for large values of the
coupling. It is not so for the coupling to $\nue$ because the latter
is bounded from above by the data on double beta decay~\cite{2-beta},
$g_{ee}< 3\cdot 10^{-5}$.  Together with the supernova bounds, the
upper limit is shifted down to $g_{ee}< 4\cdot
10^{-7}$~\cite{farzan02}, with a small window around $(2-3)\cdot
10^{-5}$. So we assume in the following that $g_{ee}< 4\cdot
10^{-7}$. For $\mu$ or $\tau$ the allowed regions are: $g_{aa} <
(3-5)\cdot 10^{-6}$ or $g_{aa}>(3- 5)\cdot 10^{-5}$. Evidently the
conditions specified above can be satisfied. As reference values we
may take $g_{ee}= 10^{-7}$ and $g_{aa}=5\cdot 10^{-6}$ which satisfy
all constraints presented above and would lead to a suppression of the
transformation of $\num$ or $\nut$ into $\nue$, thus permitting a
large muonic or tauonic asymmetry combined with a small electronic one
at BBN.  Note that it is not necessary to assume a large value of
$g_{aa} \gtrsim 10^{-5}$ to accomplish this. As shown later, such a
large coupling constant would change the scenario into something more
complex.

If majoron is lighter than the mass difference of neutrinos then a
heavier neutrino would decay into a lighter one and majoron. This
process might leave traces in the flavor ratios of the high-energy
neutrinos from distant astrophysical sources~\cite{Beacom:2002vi},
which can be detected by e.g., IceCube~\cite{Ahrens:2003ix}. This
could be potentially sensitive to very small values of the
neutrino-majoron coupling. Existing bound~\cite{bb02} on
life-time/mass of decaying neutrinos based on the solar neutrino data,
$\tau / m \geq 10^{-4}$ sec/eV, is too weak to be essential for the
mechanism discussed in the present paper.

In the subsequent sections we will present more accurate calculations
but before turning to a closer examination of the scenario, let us
make a few comments on earlier papers where majoron suppression of
neutrino oscillations in the early universe have been considered. In
Ref.~\cite{babu92} it was assumed that there exists a coupling of an
active and sterile neutrinos to majoron. Let us note that in the
version of the theory that we consider here no sterile neutrinos are
introduced: the negative helicity states are identified with
neutrinos, while the positive helicity states are identified with
antineutrinos. The coupling considered in Ref.~\cite{babu92} has the
form:
\be
{\cal L}_{as} = ig_{as}\, \chi \bar \nu_a \nu_s + h.c.
\label{l-as}
\ee
where $\nu_a$ is an active neutrino flavor and $\nu_s$ is a sterile
one. The authors argued that the effective potential induced by the
interactions of active and sterile neutrinos with majoron strongly
suppressed the oscillations. However they omitted off-diagonal
$V_{as}$-term in the effective potential which might invalidate their
result. Possibly this mechanism of oscillation suppression may operate
in a more complicated version of the model.

In Ref.~\cite{bento01} a different mechanism of neutrino oscillation
blocking by majoron has been suggested. According to the equation of
motion, the total leptonic current including majoron and neutrino
contributions is conserved:
\be
f D^2 \chi + D_\mu J^\mu = 0
\label{ddot-phi} 
\ee
where $D_\mu$ is the covariant derivative in cosmological background,
$f$ is the vacuum expectation value of the Higgs field responsible for
the spontaneous breaking of leptonic $U_L(1)$-symmetry and $J^\mu$ is
the leptonic currents of fermions. In spatially homogeneous case the
solution to this equation is
\be
\dot \chi (t) = \left[\frac {a_{in}}{a(t)}
 \right]^3\,\left( \dot\chi_{in} - J_{in}^t/f \right)  + J^t(t)/f,
\label{dot-chi}
\ee
where $a(t)$ is the cosmological scale factor and the sub-index ``in''
indicates initial values. We assume that $\dot\chi_{in}=0$.  Hence the
authors of Ref.~\cite{bento01} concluded that $\dot\chi = \eta_B
n_\gamma /f$, where $\eta_B = 6\cdot 10^{-10}$ is the baryon asymmetry
of the universe. The contribution to the neutrino effective potential
from this term is $\delta V^\chi = \dot\chi/f$ and, according
to~\cite{bento01}, with $f^2 =( 5-9)$ GeV$^2$ the potential would be
strong enough to suppress the oscillations.  However, according to the
usual scenarios, baryon asymmetry could be created at much larger
temperatures, much higher than this scale.  Thus, after spontaneous
symmetry breaking, leading to creation of majoron, $J^t$ remains
constant in comoving volume and $\dot\chi =0$. A non-zero $\dot\chi$
might be created if lepton asymmetry was generated at spontaneous
breaking of leptonic $U_L(1)$ or by the neutrino oscillations
themselves but both these cases need further and more detailed
investigation. So in our considerations we assume that the mechanism
of Ref.~\cite{bento01} is not operative.

\section{Effective potential induced by neutrino-majoron
interactions \label{s-eff-pot}}

\begin{figure}
    \centering \includegraphics[width=4cm]{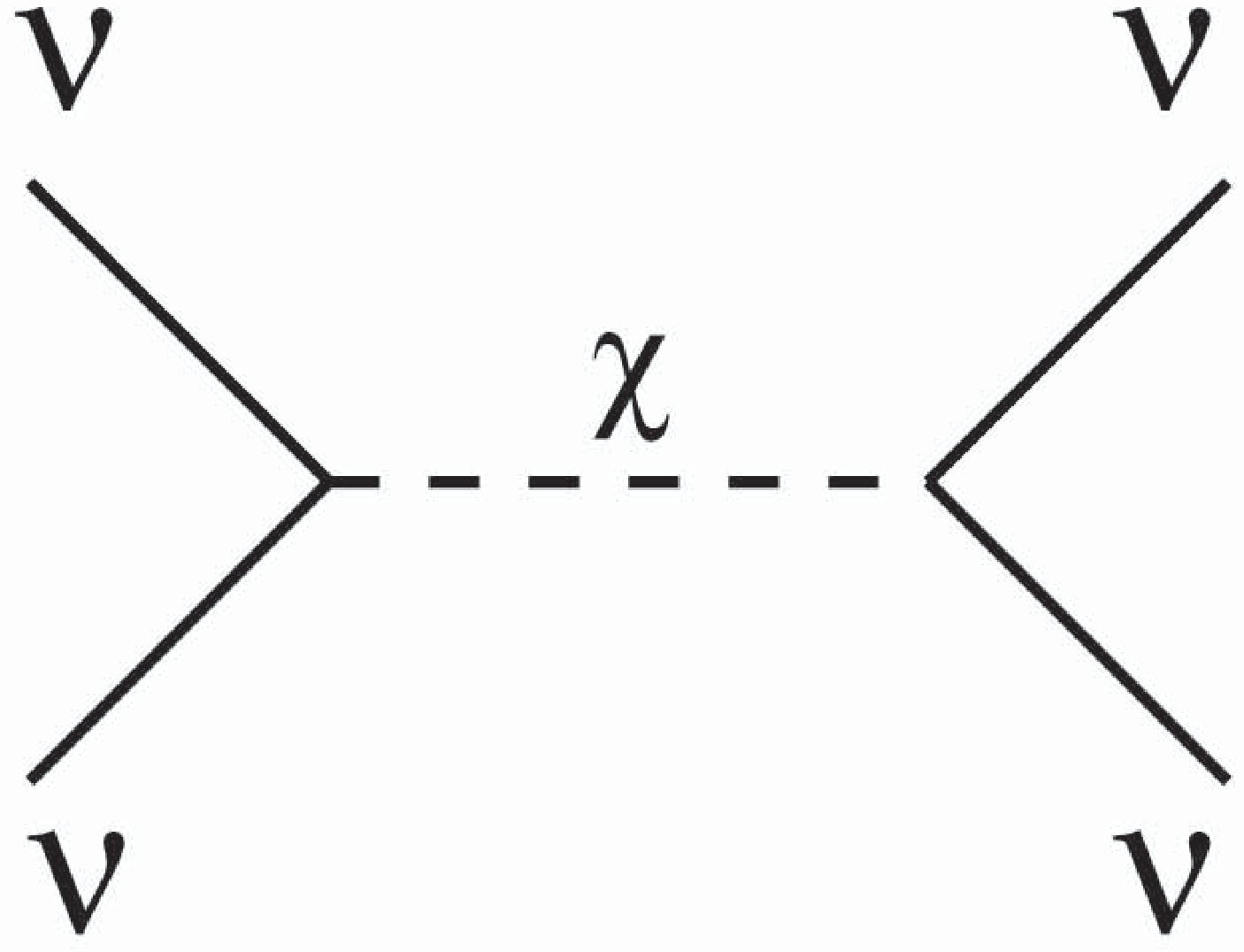}
    \includegraphics[width=4cm]{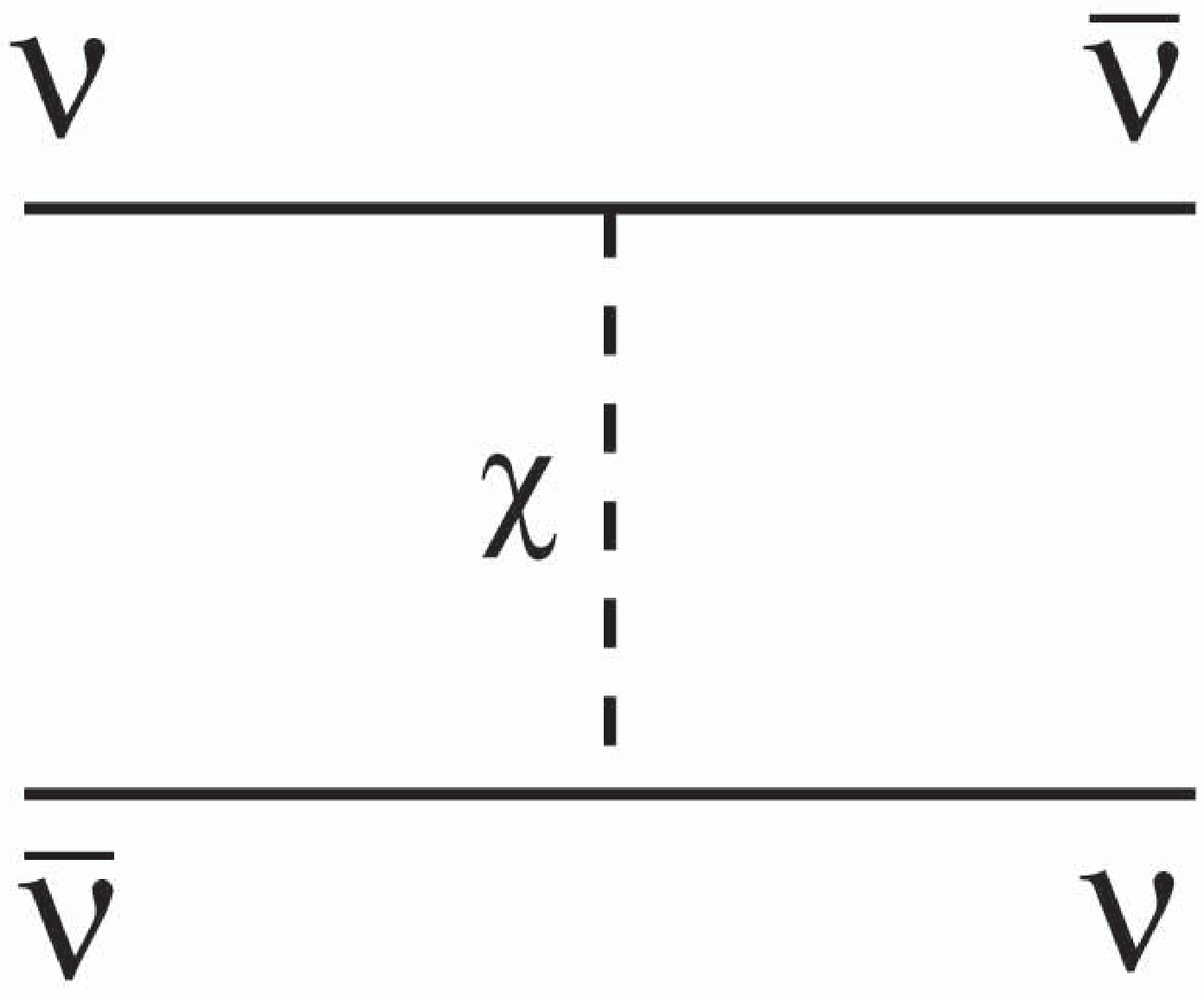} \caption{ s- and
    t-channel diagrams of neutrino-(anti)neutrino scattering through
    majoron exchange. Both diagrams contribute to the off-diagonal
    component of the effective potential.  } \label{fig:diagram1}
\end{figure}

\begin{figure}
    \centering \includegraphics[width=4cm]{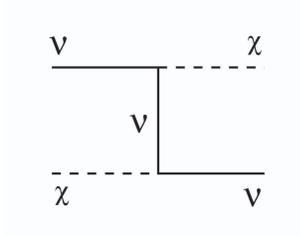} \caption{
    Diagram of neutrino-majoron scattering through neutrino
    exchange. If majorons are not abundant in the primeval plasma,
    this diagram gives negligible contribution to the effective
    potential.  } \label{fig:diagram2}
\end{figure}

In this section we derive the effective potential induced by the
neutrino-majoron interactions, which is one of essential ingredients
of our paper. The relevant part of the Lagrangian is given by
\be
{\cal L} &=&- \frac{1}{2} \partial_\mu \chi \partial^\mu \chi - \sum_a
\overline{\nu}_a \gamma^\mu \partial_\mu \nu_a + {\cal L}_{int},\\
{\cal L}_{int} &=& \frac{1}{2} \chi \left(g_{ab}\, \Phi_a ^T \sigma_2
\Phi_b +g_{ab}^* \, \Phi_b^\dagger \sigma_2 \Phi_a^* \right),\non \\
&=& \frac{i}{2} \chi \left( g_{ab}\, \nu_a ^T C \nu_b +g_{ab}^* \,
\nu_b^\dagger C \nu_a^* \right),
\label{eq:maj-nu-int}	
\ee
where $\Phi_a$ and $\nu_a$ are two-component and four-component
representations of neutrino of flavor $a$, respectively. They are
related to each other as $\nu_a^T= (\Phi_a^T,0)^T$ in the chiral
representation (see Appendix \ref{app:der} for notations).  Here and
hereafter $\nu_a$ is taken to be the left-handed field.  In the
following we neglect the effect of small masses of neutrinos, and
treat them as massless fields. In this limit, there are three diagrams
\footnote{
Note that there are other diagrams if small majorana masses are taken
into account.  However, the amplitudes of such diagrams are suppressed
by powers of $m_\nu/E_\nu
\ll 1$ for relativistic neutrinos.
} that contribute to the effective potential (see
Figs.~\ref{fig:diagram1} and \ref{fig:diagram2} ).

Let us first calculate the effective Hamiltonian corresponding to
neutrino-(anti)neutrino scattering processes. For that purpose we
integrate out the majoron field using its equation of motion, and
substitute the solution into interaction
Lagrangian~(\ref{eq:maj-nu-int}). Then we quantize neutrino fields
perturbatively. The details of the derivation can be found in the
Appendix~\ref{app:der}. The effective Hamiltonian, which describes
the neutrino-(anti)neutrino scattering processes, is given by
\be
\label{eq:eff_hamiltonian}
{\cal H}_{\rm eff}^{(\nu\nu)} (t) =
 -\frac{(2\pi)^3}{16} \int d{\bf p} \,d{\bf q}\, d{\bf r}\, d{\bf s}\,\, 
\frac { \delta^{(3)}({\bf p+q+r+s}) 
\,e^{-i E_{\rm tot} t} 
 F({\bf p}, {\bf q})F({\bf r}, {\bf s})}{\epsilon_p \epsilon_q E_{\bf p} E_{\bf q} (1-
\epsilon_p \epsilon_q
\cos \theta_{\bf pq})}
\ee
with
\be
E_{\bf p} &\equiv& \left| {\bf p} \right|,\non\\
E_{\rm tot} &\equiv& \epsilon_p E_{\bf p} +  \epsilon_q E_{\bf q} + \epsilon_r E_{\bf r} + \epsilon_s E_{\bf s} ,\non\\
\cos \theta_{\bf pq} & \equiv & \frac{{\bf p} \cdot {\bf q}}{\left|{\bf p} \right| \left|{\bf q} \right|},\non\\
F({\bf p}, {\bf q})& \equiv &  g_{ab}\, \nu_a({\bf p}) ^T C \nu_b({\bf q}) 
+g_{ab}^* \, \nu_b({\bf -q})^\dagger C \nu_a^*({\bf -p}),
\ee
where $d{\bf p}  \equiv  d^3 {\bf p}/(2 \pi)^3$.
The  momentum expansion of free neutrino field is
\be
\nu_i(x) &=& \int d{\bf p}~ \nu_i({\bf p})\, e^{-i p^0 t +i {\bf p} \,{\bf x}},\non\\
&=& \int d{\bf p}~\left(a_i({\bf p}) u_{\bf p} + b_i({\bf -p})^\dagger v_{\bf- p}  \right)
 \, e^{-i p^0 t +i {\bf p} \,{\bf x}}
\ee
with $p^0  \equiv \epsilon_p \left|{\bf p} \right|= 
\pm  \left|{\bf p} \right|$. The sign of $\epsilon_p$
is chosen to reproduce positive or negative energy solution.  Here
$a_i({\bf p})$ ($b_i{}^\dagger({\bf p})$) is the annihilation
(creation) operator for negative (positive)-helicity neutrinos with
momentum ${\bf p}$, $u_{\bf p}$($v_{\bf p}$) represents left-handed
Dirac spinor of negative (positive)-helicity neutrinos. With thus
obtained effective Hamiltonian, the effective potential can be
calculated by the technique of Sigl and Raffelt~\cite{Sigl:fn}.
Although the calculations are lengthy, the procedure is
straightforward. The contribution to the effective potential for
neutrinos with momentum ${\bf p}$ is given by
\be
\label{eq:eff_potential}
\left[V_{\bf p}^{(\nu\nu)}\right]_{ab} = \int d{\bf q}~~\frac{1}{4 \left|{\bf p} \right| \left|{\bf q} \right|} \left[
g^\dagger \left(\rho^T_{\bf q} + \overline{\rho}^T_{\bf q} \right) g
\right]_{ab},
\ee
where the density matrices are defined as
\be
\la a^\dagger_j({\bf p}) \,a_i({\bf p}')\ra &=& (2 \pi)^3 \delta^{(3)}({\bf p}-{\bf p}') 
\left[\rho_{\bf p} \right]_{i j},\non\\
\la b^\dagger_i({\bf p}) \,b_j({\bf p}')\ra &=& (2 \pi)^3 \delta^{(3)}({\bf p}-{\bf p}') 
\left[\overline{\rho}_{\bf p} \right]_{i j}.
\ee
If we take the coupling constant matrix $g_{ab}$ to be diagonal, the
effective potential becomes
\be
\label{eq:eff_potential_diag}
\left[V_{\bf p}^{(\nu\nu)}\right]_{ab} &=& g^*_{aa} g_{bb} \int d{\bf q}~~\frac{1}{4 \left|{\bf p} 
\right| \left|{\bf q} \right|} 
 \left(\rho^T_{\bf q} + \overline{\rho}^T_{\bf q} \right)_{ab}
 ~~~(\rm{no~summation})
 ,\non\\
 &\sim& g^*_{aa} g_{bb} \frac{T^2}{\left|{\bf p} \right|}.\non
\ee
This result confirms the rough estimate of the effective potential in
the previous section.

The neutrino-majoron scattering shown in Fig.~\ref{fig:diagram2} also
contributes to the effective potential and can be calculated
similarly. The result is
\be
\label{eq:eff_potential2}
\left[V_{\bf p}^{(\nu\chi)}\right]_{ab} = \int d{\bf q}~~\frac{f_{\chi}({\bf q})}{
4 \left|{\bf p} \right| \left|{\bf q} \right|} \left[ g^\dagger g\right]_{ab},
\ee
where  $f_{\chi}({\bf p})$ is the number density of 
majorons with momentum $\bf{p}$.
Note that this expression vanishes if the abundance of majorons is 
negligible in comparison with their equilibrium value,
i.e. $f_{\chi}({\bf p}) \ll f_\chi^{(eq)}({\bf{p}})$. Thus the complete 
effective 
potential for neutrinos with momentum ${\bf p}$ induced by interactions
with majorons is  given by
\be
\left[V_{\bf p}^{(\chi)}\right]_{ab} = \int d{\bf q}~~\frac{1}{4 \left|{\bf p} \right| \left|{\bf q} \right|} \left[
g^\dagger \left(\rho^T_{\bf q} + \overline{\rho}^T_{\bf q} 
+f_{\chi}({\bf q})\cdot {\bf 1}
\right) g
\right]_{ab},
\label{eq:eff_potential_total}
\ee
where ${\bf 1}$ is the unit matrix in the flavor basis.


\section{Possible role of neutrino-majoron reactions. \label{s-react}}

\begin{figure}
    \centering
    \includegraphics[width=4cm]{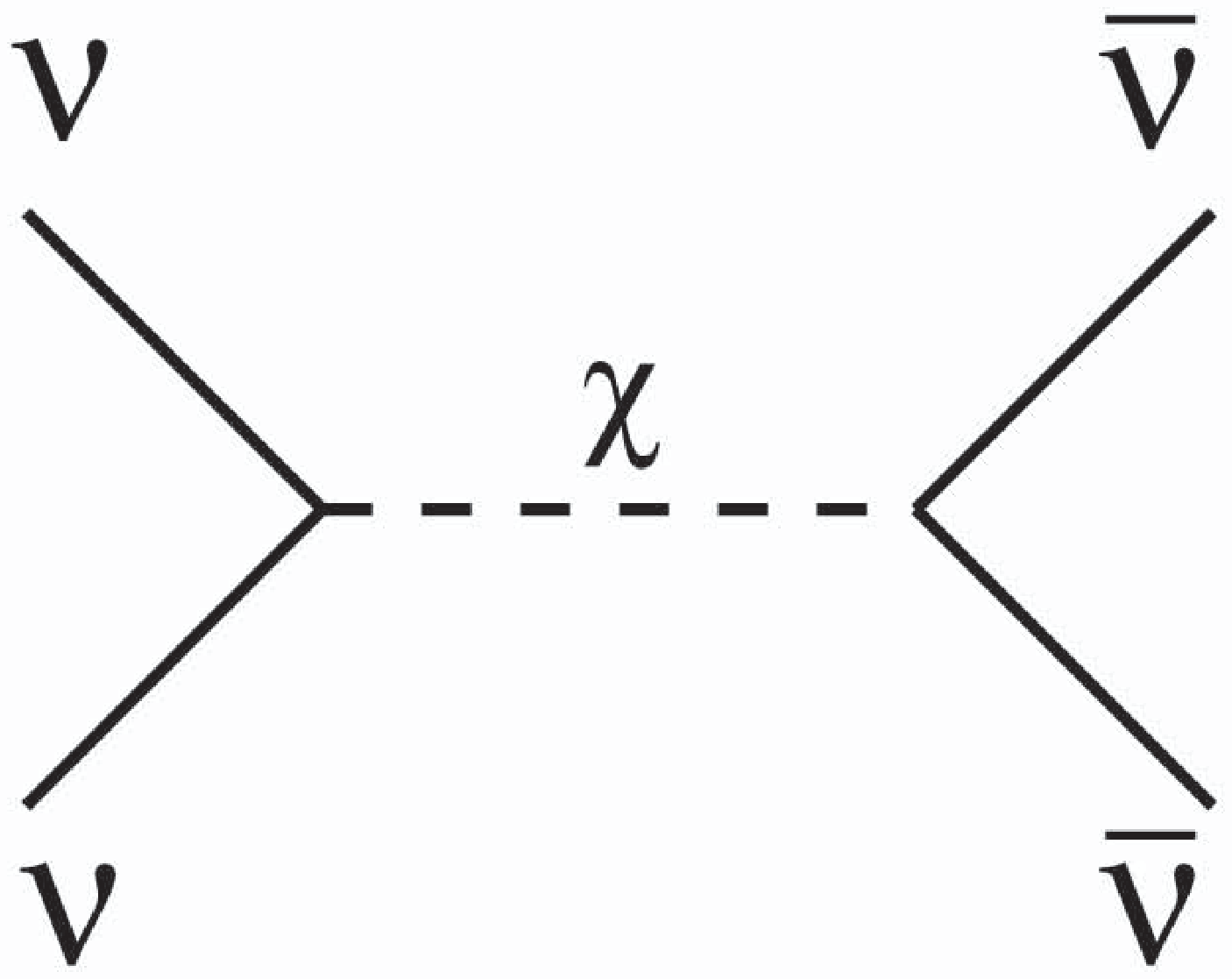}
    \includegraphics[width=4cm]{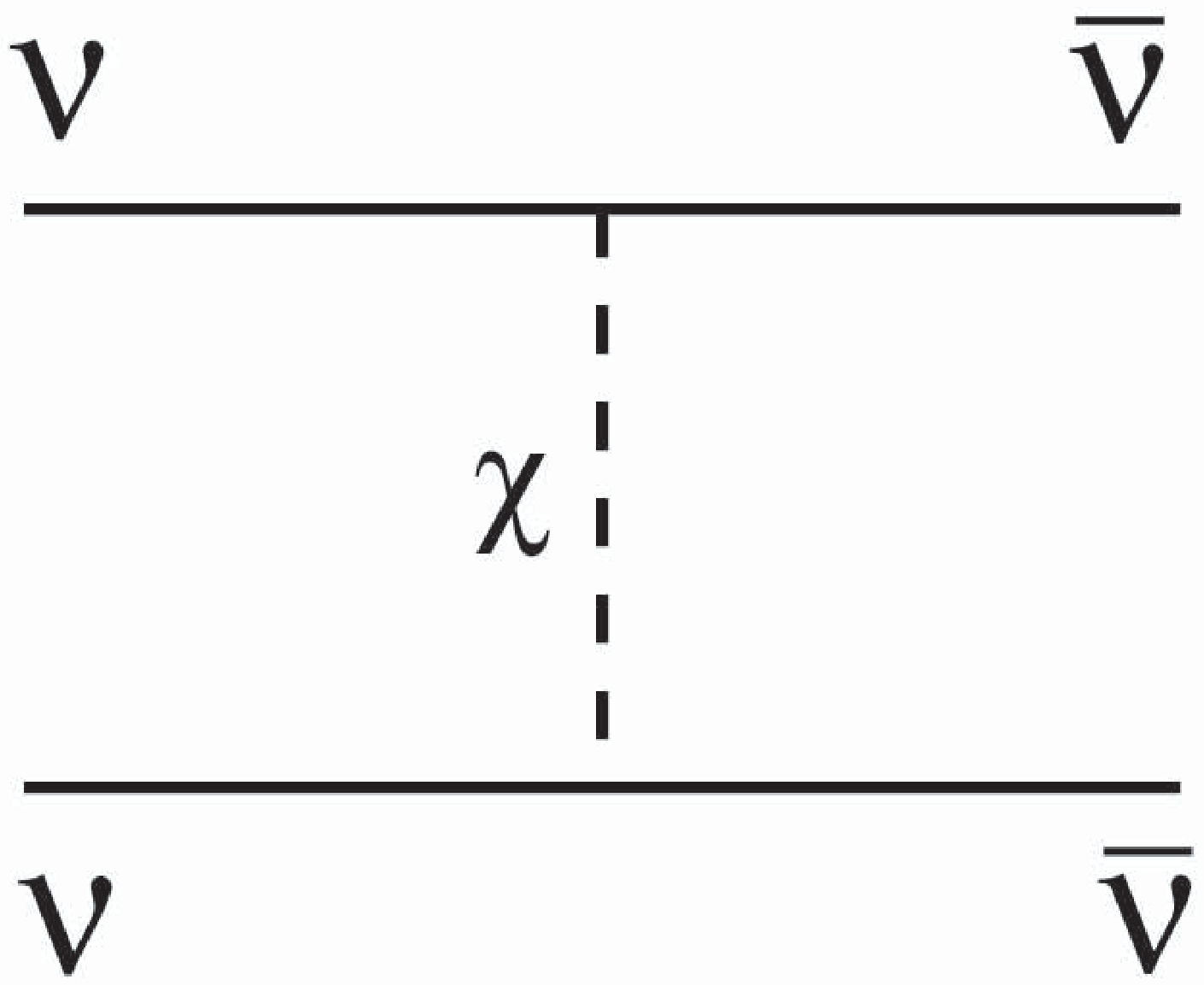}
    \includegraphics[width=4cm]{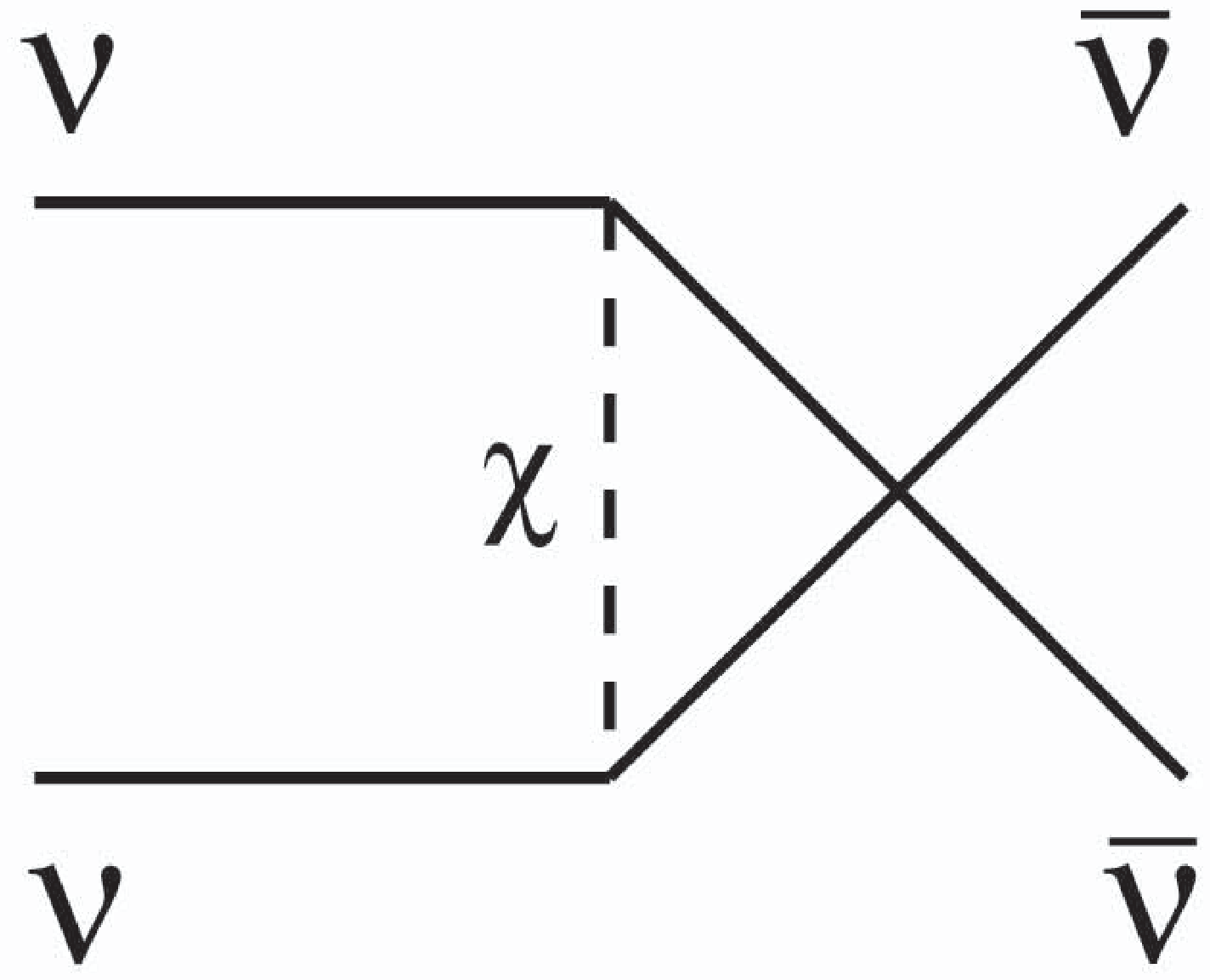}
 \caption{
s-, t- and u-channel diagrams of $2 \nu \leftrightarrow 2 \bar\nu$ through majoron
exchange. 
    }
    \label{fig:diagram3}
\end{figure}

As we have already mentioned reactions between neutrinos with an
exchange of majorons do not conserve leptonic charge and if they are
efficient, lepton asymmetry of the universe would be completely
destroyed before BBN. One of the dangerous processes is
\be
2\nu \lrar 2\bar\nu.
\label{2nu-2barnu}
\ee
We assume for simplicity that the diagonal coupling constants
$|g_{aa}|$ are larger than non-diagonal ones and that one flavor
coupling dominates, e.g.  $|g_{\tau\tau}|>|g_{\mu\mu}|>|g_{ee}|$. In
this case only three diagrams presented in Fig.~\ref{fig:diagram3}
contribute into the reaction (\ref{2nu-2barnu}).  The amplitude
corresponding to any of these three diagram is equal to $|g|^2$ and
thus the complete amplitude squared is simply $|A(2\nu\lrar
2\bar\nu)|^2 = 9|g|^4$ (see Appendix~\ref{app:amplitude}).  Here and
in what follows we omit the sub-index $aa$ at the coupling constant
$g_{aa}$.  With the known amplitude of the $L$-nonconserving reaction
($L$ is the leptonic charge), the kinetic equation can be written as
\be
 \frac{h}{x^2}\,x \,\partial_x f_{\bar\nu} (y_1) 
= \frac{1}{2} \cdot \frac{9 |g|^4}{2^{9}\pi^5xy_1}
\int \frac{d^3 \bf{y}_2}{y_2}
\frac{d^3 {\bf y}_3}{y_3} \frac{d^3{\bf y}_4}{y_4}\,
F(f)\,\delta^{(4)} (y_1 + y_2 -y_3 -y_4) 
+ I_{coll}^{(el)}
\label{f-barnu}
\ee
where $f$ is the neutrino distribution function, $(y_i,{\bf
y}_i)=(E_i/T,{\bf p}_i/T)$, $x=(T/{\rm MeV})^{-1}$, $T$ is the
temperature of the cosmic plasma, $(H/ {\rm MeV}) = h/x^2$ is the
Hubble parameter with $ h \approx 4.5\cdot 10^{-22}$,
$I_{coll}^{(el)}$ is the contribution to the collision integral from
reactions that conserve leptonic charge, and
\be
F(f) = [ 1 - f_{\bar\nu} (y_1)][ 1 - f_{\bar\nu} (y_2)]
f_{\nu} (y_3) f_{\nu} (y_4)
- [ 1 - f_{\nu} (y_3)][ 1 - f_{\nu} (y_4)]
f_{\bar\nu} (y_1) f_{\bar\nu} (y_2).
\label{F}
\ee
We assumed that the temperature evolved according to $\dot T = -HT$.
The combinatorial factor $1/2$ in front of the r.h.s. comes from two
factors $1/2!$ and one factor 2 because two identical particle are
produced (see e.g. Ref.~\cite{adS}).

\begin{figure}
    \centering \includegraphics[width=10cm]{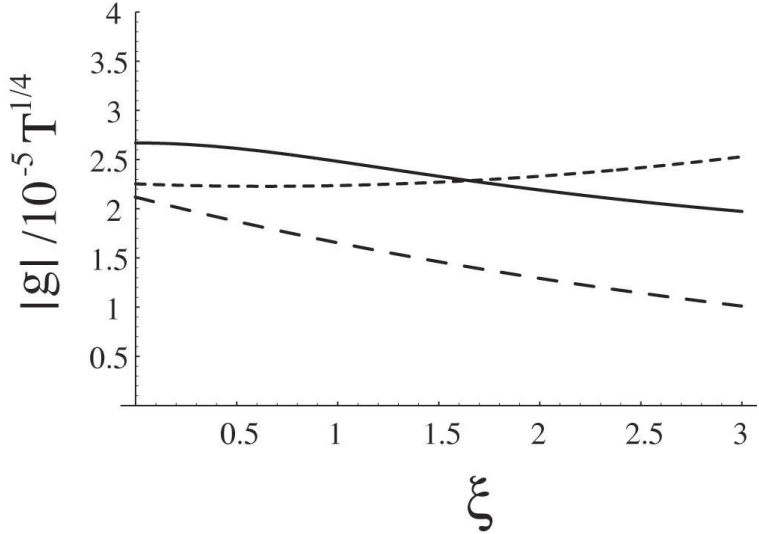} \caption{
    Upper bounds on $|g|$ obtained from the requirement that
    $L$-violating reactions, $2 \nu \leftrightarrow 2 \bar\nu$ and
    $\nu+\bar \nu \leftrightarrow 2 \chi$, are out of
    equilibrium. Numerical and analytical bounds for the former
    reaction are shown as solid and long-dashed lines,
    respectively. The short-dashed line represents the constraint
    obtained numerically for the latter reaction.  }
    \label{fig:g-fermi}
\end{figure}

The collision integral in Eq.~(\ref{f-barnu}) can be easily evaluated
in the limit of Boltzmann statistics, when $f \ll 1$ and under
assumption of kinetic equilibrium, so the distribution functions have
the form $f= \exp (-y + \xi)$, where $\xi = \mu/T$ is the
dimensionless chemical potential. To avoid the contribution of
$I_{coll}^{(el)}$ to the equation governing the evolution of $\xi$ we
take the difference of Eq.~(\ref{f-barnu}) and analogous equation for
$f_\nu$ and integrate it over $d^3 {\bf y}_1$. The contribution of
$I_{coll}^{(el)}$ to this integrated difference vanishes. The
resulting equation takes the form:
\be
\partial_x \left( e^\xi-e^{\bar\xi}\right) = 
-\frac{9|g|^4}{2^7\pi^3 h}\left(e^{2\xi} -e^{2\bar\xi}\right),
\label{d-xi}
\ee
where we have used the formula presented in
Appendix~\ref{app:formula}.  It is also assumed above that $\xi=-\bar
\xi$. Otherwise the contribution from $L$-conserving $\bar\nu
\nu$-annihilation would not vanish in the collision integral. If
$\xi\geq 1$ we may neglect $\exp (-\xi)$ and solve equation
(\ref{d-xi}) as
\be
\exp [\xi (x)] = \frac{\exp [\xi (0)]}{1 + 9|g|^4 x \exp[\xi(0)]/(128\pi^3 h)}
\label{xi-of-x}
\ee
The lepton asymmetry would not be destroyed by reaction 
(\ref{2nu-2barnu}) if 
\be
|g| < 2\cdot 10^{-5} T^{1/4}\exp [-\xi (0)/4] .
\label{g-asym}
\ee
Here (and in what follows) $T$ is given in MeV.  Quantum statistics
(Fermi) effects somewhat weaken the bound. Their presence does not
allow to solve kinetic equation analytically. Numerical estimates give
a similar constraint as shown in Fig.~\ref{fig:g-fermi}.

Muonic or tauonic charge asymmetry should be preserved until the
annihilation $\bar \nu_{\tau,\mu}\nu_{\tau,\mu}$ into $e^+e^-$-pairs
is frozen. Otherwise we will return to the standard scenario with zero
lepton asymmetry.  According to the estimates of Ref.~\cite{ad-itep02}
made in Boltzmann approximation, the freezing temperature of the
annihilation is $T_f\approx 5.3$ MeV.  Fermi corrections would make
its value slightly higher.  If muonic and/or tauonic charge asymmetry
were erased by the oscillations or L-nonconserving reactions with
exchange of majorons below $T_f$ then the total number density of
$\nut$ plus $\bar\nut$ (or $\num +\bar\num$) would be conserved in the
comoving volume and their distribution would be given by
\be
f_{\nu_a} = f_{\bar\nu_a} = \left[ \exp (E/T - \xi_a )\right]^{-1}
\label{fnu=fbar}
\ee
where $a=\mu,\tau$.  So effectively the energy density of these
neutrinos would be the same (up to a numerical factor of order of
unity) as the energy density of the usual degenerate neutrinos. If the
mixing of $\nut$ or $\num$ with $\nue$ would not change the
distribution of the latter then BBN would allow a large values of
$\xi_a$ and the degeneracy of $\nu_a$ may lead to noticeable
cosmological effects.

\begin{figure}
    \centering \includegraphics[width=4cm]{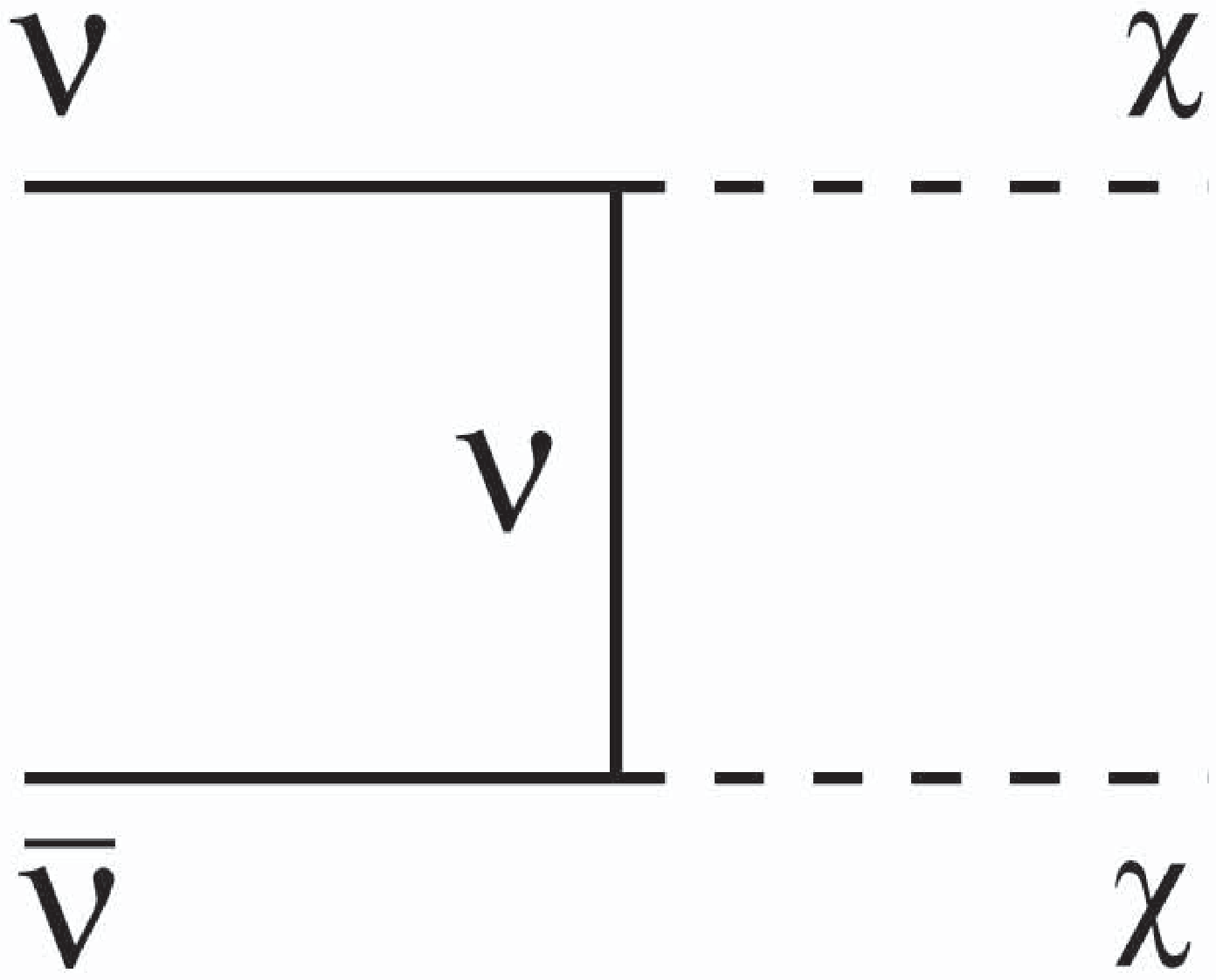}
    \includegraphics[width=4cm]{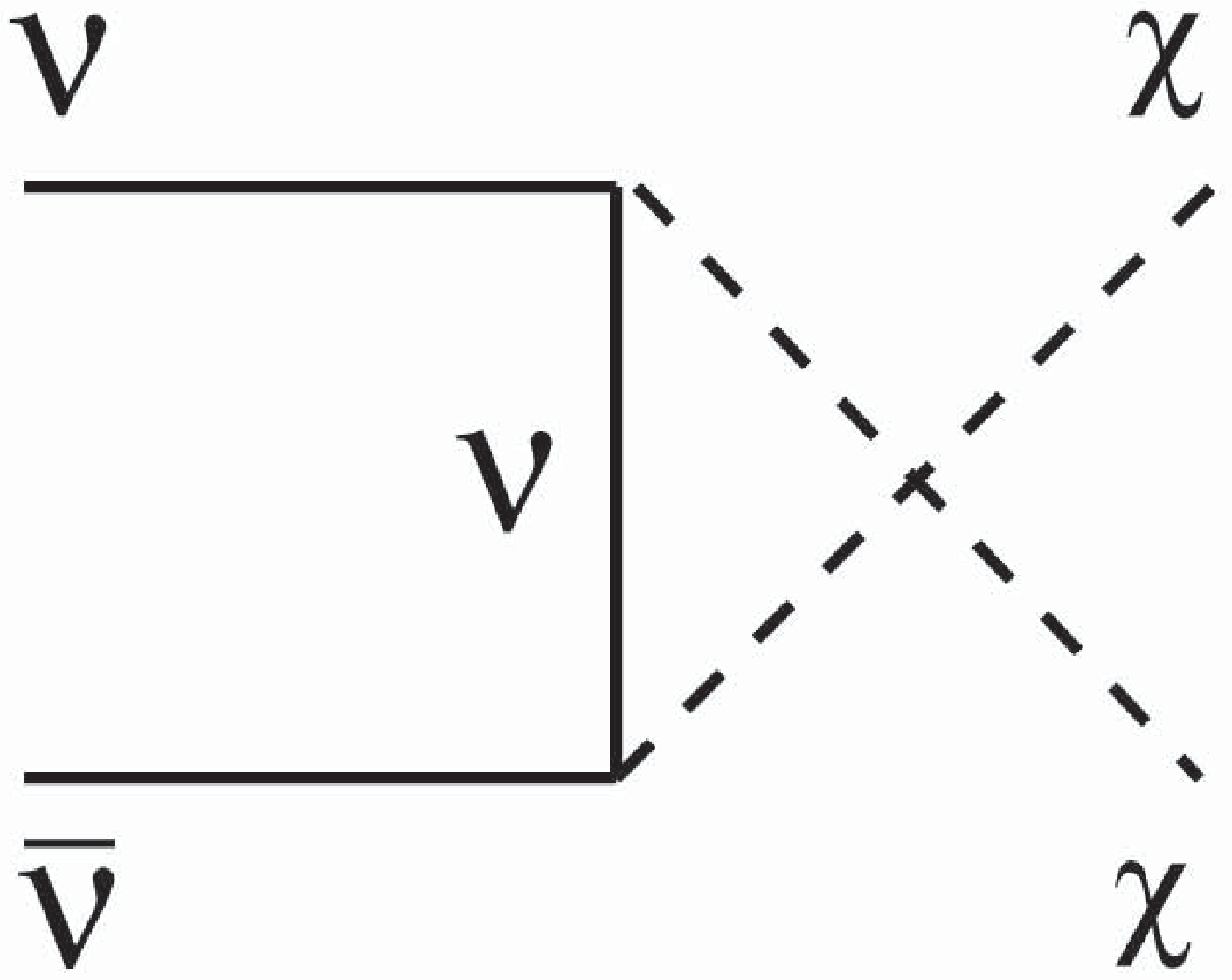} \caption{ t- and
    u-channel diagrams of $ \nu+\bar\nu \leftrightarrow 2 \chi$
    through neutrino exchange.  } \label{fig:diagram4}
\end{figure}

We make a simplifying assumption of absence of majorons in noticeable
amount in the primeval plasma. If majorons would be in equilibrium we
should take into account their (4/7)-contribution into the number of
effective neutrino species at BBN and modification of neutrino
effective potential due to forward elastic scattering $\nu \chi \rar
\nu \chi$.  Even if majorons are abundantly produced, the main
conclusion of our paper remains unchanged but numerical results would
be somewhat different.  The production of majorons could proceed
through the reaction $\nu +\bar\nu \rar 2\chi$.  The Feynman diagrams
describing this process are presented in Fig.~\ref{fig:diagram4}.  The
amplitude squared of this process is $$\left|M_{\rm
inv}\right|^2=|g|^4 \left(\frac{p_1 \cdot p_3}{p_1 \cdot p_4}
+\frac{p_1 \cdot p_4}{p_1 \cdot p_3}-2\right),$$ where $p_{1(2)}$ and
$p_{3,4}$ are four momenta for (anti)neutrino in the initial state and
majorons in the final state, and they satisfy the momentum
conservation condition, $p_1+p_2 = p_3+p_4$. Note that the collision
integral with this amplitude squared involves an IR logarithmic
divergence, therefore we need to input a lower cutoff scale. Taking
into account the thermal effects, the dispersion relations for the
majoron and the neutrinos change from those at zero
temperature. Especially, they obtain finite thermal masses, which
provide the desired infra-red cutoff.  For $|g| \sim O(10^{-5})$, the
thermal effect of the neutrino-majoron interaction dominates over that
due to the electroweak interaction, so the IR divergent part is
regularized as $\sim \log(4E^2/m_{T}^2)$ with $m_{T}
\sim |g| T$.

Kinetic equation governing the production of majorons is similar to
Eq.~(\ref{f-barnu}), and has the form:
\be
\frac{h}{x^2} x\partial_x f_\chi = \frac{|g|^4\log(4E^2/m_{T}^2)}{2^7 \pi^3 x}\,\frac{\exp (-y)}{y}+... 
\label{f-chi}
\ee
where we omitted elastic reactions. Integrating this equation we can
find for the ratio of the number density of $\chi$ to the equilibrium
value:
\be
\frac{n_\chi}{n_\chi^{(eq)}} = \frac{|g|^4 \log(4E^2/m_{T}^2) ~x }{2^8 \zeta(3) \pi^3 h}
\label{chi/eq}
\ee
Demanding that $n_\chi <n_\chi^{(eq)}$ we obtain
\be
|g| < 2.1\cdot 10^{-5} T^{1/4},
\label{g-prod}
\ee
which gives a constra
int similar to Eq.~(\ref{g-asym}). However,
if we take into account quantum statistics, the bound becomes 
slightly weaker for 
large charge asymmetry of neutrinos, as shown in Fig.~\ref{fig:g-fermi}. 

So, to summarize, the muonic or tauonic lepton asymmetry would be
erased if the neutrino-majoron interaction is sufficiently strong,
i.e. $|g| \gtrsim 10^{-5}$. If we restrict ourselves to a scenario
that the majoron-neutrino interactions suppress neutrino oscillations
and thereby keep the large lepton asymmetries intact, the largest
coupling constant should be smaller than $\sim 10^{-5}$. However this
does not necessarily mean that the conspiracy between the speed-up
effect of $\xi_{\mu,\tau}$ and a shift of $\beta$ equilibrium due to
$\xi_{e}$ is impossible for $|g| \gtrsim 10^{-5}$. In fact, if the
$L$-nonconserving interactions became efficient only after muonic and
tauonic neutrinos decoupled from the thermal plasma, then the neutrino
degeneracy would still be maintained in muonic or tauonic neutrinos
resulting in their larger energy density at BBN.  Majorons, as well,
can contribute to additional relativistic degrees of freedom at BBN.
We will discuss these issues below.


\section{Neutrino oscillations in primeval plasma \label{s-kin-eq}}

As is established now, neutrino mass eigenstates are related to the
flavor eigenstates through the orthogonal matrix (we neglect a
possible CP violation):
\be
\nu_a = U_{a j} \,\nu_j
\label{nuj-nua}
\ee
where $a = e,\mu,\tau$ and $j=1,2,3$. We have chosen the parameters so
that in the limit of small mixing $\nue \approx \nu_1$, $\num \approx
\nu_2$, $\nut\approx \nu_3$. However, the mixing between neutrinos are
known to be large and they cannot be taken as a dominant single mass
eigenstate.

Kinetic equations for the density matrix of oscillating
neutrinos can be written as~\cite{kin-eq,Sigl:fn}:
\be
i\dot \rho = [{\cal H}^{(1)},\rho ] - i \{{\cal H}^{(2)},\rho \} 
\label{dotrho}
\ee
where the first commutator term includes the vacuum Hamiltonian and
the effective potential of neutrinos in medium calculated in the first
order in Fermi coupling constant, $G_F$ (weak interaction part) and in
the second order in the coupling to majoron. The latter is given by
Eq.~(\ref{eq:eff_potential_total}) and does not contain off-diagonal
terms if only one flavor coupling constant dominates. The weak
interaction potential contains off-diagonal terms of the same order of
magnitude as diagonal ones because of universal coupling of $W$ and
$Z$ bosons to all neutrino flavors.  Their explicit expressions can be
found e.g. in Ref.~\cite{act-act}.  In what follows we will skip upper
indices $(1)$ and $(2)$ at ${\cal H}$.

The second anti-commutator term in Eq.~(\ref{dotrho}) describes
breaking of coherence induced by neutrino scattering and annihilation
as well as neutrino production by collisions in primeval plasma. It
includes imaginary part of the Hamiltonian calculated in the second
order in $G_F$ and the fourth order terms in neutrino-majoron coupling
$g$ related to neutrino-neutrino scattering through majoron
exchange. If majorons were abundant in the primeval plasma the
processes of neutrino-majoron scattering should also be included.

Since the ``atmospheric'' neutrino mass difference is much larger than
the ``solar'' one, we may simplify the problem considering the process
in two steps. First, oscillations between $\nut$ and $\num$ were
switched-on. This started at temperatures about 10-15 MeV and might
lead to equilibration of tauonic and muonic charge asymmetries if
these asymmetries were not erased by reactions
(\ref{2nu-2barnu}). Later at $T\leq 3$ MeV, oscillations would start
between $\nue$ and $\nu_{\mu'}$, which is a certain mixture of $c\nut
+ s \num$. This process is potentially dangerous for BBN because it
could change number density and spectrum of $\nue$ and $\bar\nue$.

If mixing is effective only between two neutrinos, the density matrix is
$2\times 2$ and kinetic equations for its components have the form:
\be
iHx\partial_x \rmm &=& \hmt\rtm - \htm\rmt - i I^{(coll)}_{\mu\mu}
\label{drmm}
\\
iHx\partial_x \rtt &=& \htm\rmt - \hmt\rtm - i I^{(coll)}_{\tau\tau}
\label{drtt}
\\
iHx\partial_x \rmt &=& \left( \hmm -\htt\right)\rmt +
\hmt \left(\rtt -\rmm \right) - i\Gamma_{\mu\tau} \rmt
\label{drmt}
\ee
where $H$ and $x$ are defined below Eq.~(\ref{f-barnu}), $\hab =
V_{ab} + \sum_{j=2,3} U_{aj}U_{bj} m^2_j/2E$, $U_{aj}$ are matrix
elements of the mixing matrix; in the case under consideration $U_{\mu
2} = U_{\tau 3} =\cos \theta \equiv c$ and $U_{\mu 3} = -U_{\tau 2}
=\sin \theta \equiv s$.  The coherence breaking terms are given by the
usual collision integrals $I^{(coll)}$ in the equations for the
diagonal components and by
\be
\Gamma_{\mu\tau} = 1.1\cdot 10^{-22} (y/x^5) \,{\rm MeV}
\label{gam-mu-tau}
\ee
for the non-diagonal component $\rmt= \rtm^*$. In the expression for
$\Gamma_{\mu \tau}$ we neglected the contribution from the majoron
related processes. The effective potential $V_{ab}$ contains
contributions from the usual weak interactions and from the
neutrino-majoron interactions (\ref{eq:eff_potential_total}).  We
assume that the weak part is dominated by the charge asymmetric
contribution, which is true in the case of a large charge asymmetry:
\be
V^{(w)}_{ab} = 1.5\, C_1\, G_F T^3 \int d{\bf y} \left(\rho_{ab} - \bar\rho_{ab}
\right) \equiv B \int d{\bf y} \left(\rho_{ab} - \bar\rho_{ab}
\right). 
\label{vw-ab}
\ee
with $C_1$ defined in Eq.~(\ref{veff}), $d{\bf y} = d^3 {\bf
y}/(2\pi)^3$, $\bar\rho$ is the antineutrino density matrix, and the
coefficient 1.5 comes from $(11/4) \cdot 2\, \zeta(3)/\pi^2$,
i.e. from normalization to photon number density.

Equations (\ref{drmm}-\ref{drmt}) can be solved numerically but we can
make reasonable estimates analytically in the following way (for more details
see e.g. Refs.~\cite{dolgov02,ad-fv-03}). In the case of strong coherence
breaking, i.e. $\Gamma_{\mu\tau} \gg H$,  Eq.~(\ref{drmt}) can be formally solved
in the stationary point approximation i.e. putting r.h.s. equal zero. In this
way we obtain:
\be
\rmt = 
\frac{\hmt \left(\rmm - \rtt \right)}{\hmm - \htt -i\Gamma_{\mu\tau}} 
\approx
\frac{ \rmm - \rtt }{\vmm - \vtt}\,\left( V_{\mu\tau} + \frac{\delta m_{32}^2}{4E}\,s_2
\right)\,
\left(1 + \frac{i\Gamma_{\mu\tau} - c_2 \delta m_{32}^2/2E}{\vmm -\vtt}\right) 
\label{rmt}
\ee
where $\delta m_{32}^2 \equiv m_3^2-m_2^2$,
$s_2 = \sin 2\theta$ and $c_2 = \cos 2\theta$.

This expression could be substituted into Eqs.~(\ref{drmm}) and
(\ref{drtt}) for the diagonal components but one cannot obtain the
closed equations for the latter because the potential $\vmt$ contains
an integral from $\rmt$ over momentum, see Eq.~(\ref{vw-ab}). We
assume for simplicity that only weak-interaction potential has
noticeable off-diagonal components, i.e. $V_{\mu\tau}\simeq
V_{\mu\tau}^{(w)}$. It can be true if e.g. $|g_{\tau\tau}| \gg
|g_{\mu\mu}|$. To express $\vmt$ through diagonal components we
integrate Eq. (\ref{rmt}) and similar equation for $\bar\rmt$ over
momentum and subtract one from the other.  To simplify the expressions
let us present the diagonal part of the potential as a sum of charge
symmetric (majoron) and antisymmetric (weak) parts:
\be
\vmm -\vtt \equiv V_+^{(\chi)} + V_-^{(w)} 
\label{vmm-vtt}
\ee
where the charge asymmetric part $V_-^{(w)}$ can be separated into two
terms containing integrals from neutrino and antineutrino elements of
the density matrix (see Eq.~(\ref{vw-ab})), $V_-^{(w)} = V_\nu - \bar
V_\nu$.

After straightforward calculations we obtain:
\be
\vmt = \frac{ V_+^{(\chi) \,2} - (V_\nu - \bar V_\nu)^2}
{ V_+^{(\chi)} ( V_+^{(\chi)}- V_\nu - \bar V_\nu)}
 \int d{\bf y}\, \frac{s_2 \delta m_{32}^2 B }{4E}\left[ \frac{ \rmm -\rtt} 
{V_+^{(\chi)}+ V_\nu - \bar V_\nu}
\left( 1+\frac{i\Gamma_{\mu\tau}-\delta m_{32}^2 c_2 /2E}{V_+^{(\chi)}+ V_\nu - \bar V_\nu}
\right) \right. 
\nonumber \\
\left.  - \frac{ \bar\rmm -\bar\rtt} 
{V_+^{(\chi)}- V_\nu + \bar V_\nu} 
\left( 1+\frac{i\bar\Gamma_{\mu\tau}-\delta m_{32}^2 c_2 /2E}
{V_+^{(\chi)}-V_\nu+\bar V_\nu} \right)
\right].
\label{vmt}
\ee
This result is valid if $|V_+^{(\chi)}| \geq |V_-^{(w)}|$. In the case
of the usual weak interactions considered e.g. in
Refs.~\cite{act-act,lunardini01} the situation is opposite,
$|V_+^{(\chi)}| \ll |V_-^{(w)}|$, and the character of the approximate
analytical solution is completely different.

For a rough estimate let us assume that $|V_+^{(\chi)}| > |V_-^{(w)}|$
and $|V_+^{(\chi)}| > |\delta m_{32}^2|/2E$, while $|\vmt| < |s_2
\delta m_{32}^2| /2E$, as follows from Eq.~(\ref{vmt}). In this case
we find
\be
Hx\partial_x \rmm = -\left(\frac{\delta m_{32}^2}{4E}\, s_2 \right)^2 \,
\frac{\Gamma_{\mu\tau} (\rmm - \rtt)}{(V_{\mu\mu}^{(\chi)}-
V_{\tau\tau}^{(\chi)})^2}
- I^{(coll)}_{\mu\mu}.
\label{drmm2}
\ee
The evolution of the difference of muonic and tauonic charges is
determined by
\be
L^{(\mu\tau)}&\equiv&\int d{\bf y}\, L_{\bf y}^{(\mu\tau)} \equiv
\int d{\bf y} \left(\rmm - \rtt - \bar\rmm + \bar\rtt \right).
\label{L}
\ee
The collision integrals disappear from the time derivative of this difference 
since weak interactions
conserve leptonic charges and majoron induced processes are assumed to be
suppressed, see sec.~\ref{s-react} and in particular Eq.~(\ref{g-asym}).
The evolution of $L$ is governed by the equation:
\be
\frac{dL^{(\mu\tau)}}{dx} \approx 
\left(\frac{|g|}{1.5\times10^{-7}}\right)^{-4} 
\left(\frac{\delta m_{32}^2}{2.5\times10^{-3}{\rm eV}^2}\right)^{2}  
\int \frac{d{\bf y}}{y} \,L_{\bf y}^{(\mu\tau)}
\label{dL/dx}
\ee
and for the majoron coupling constant bounded by the conditions
presented above, $L^{(\mu\tau)}$ remains practically constant till
nucleosynthesis epoch, i.e.  $x\sim 1$.

\begin{figure}
    \centering
    \includegraphics[width=10cm]{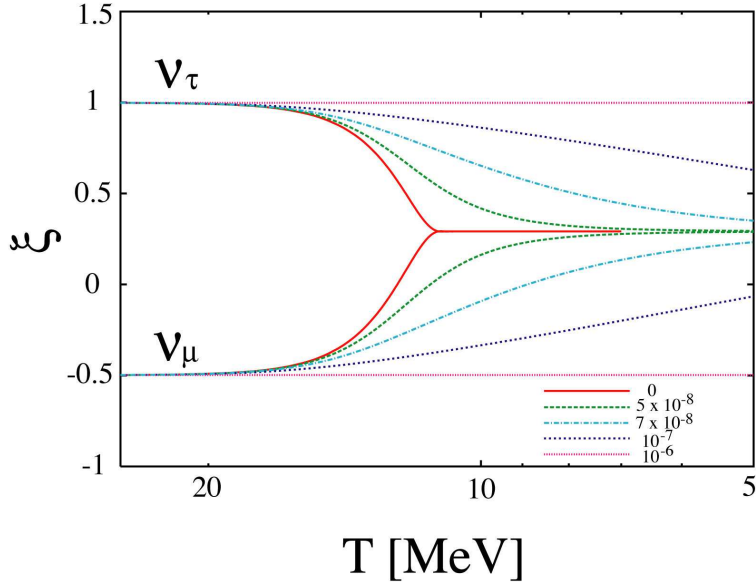}
    \caption{
The evolutions of $\xi_\mu$ and $\xi_\tau$ for several values of $|g|$
for the maximal mixing and $\delta m_{32}^2 = 2.5 \times 10^{-3} {\rm eV}^2$. 
The initial conditions are $\xi_\mu = -0.5$ and $\xi_\tau=1.0$.
    }
       \label{fig:diagram6}
\end{figure}

We have performed numerical calculations to solve
Eqs.~(\ref{drmm}-\ref{drmt}) in a similar way to Ref.~\cite{act-act}.
The coupling constant matrix $g_{ab}$ is assumed to take the form:
$g_{ab} = g \,\delta_{a \tau} \delta_{b \tau}$, for simplicity.  The
evolution of dimensionless chemical potentials $\xi_\mu$ and
$\xi_\tau$ for several values of $|g|$ are shown in
Fig.~\ref{fig:diagram6}.  One can see that the lepton asymmetries of
$\nu_\mu$ and $\nu_\tau$ equilibrate around $T \sim 12$MeV when
$|g|=0$. We have checked that the evolution remains the same for $|g|
< 10^{-8}$. As $|g|$ increases, the oscillations become less efficient
and completely stop for $|g| \gtrsim 10^{-7}$, which well agrees with
the analytical estimates. It should be also noted that the lower bound
on $|g|$ obtained from Eq.~(\ref{dL/dx}) does not depend on the
magnitude of the initial lepton asymmetries. We have checked that the
oscillations are similarly stuck for $|g| \gtrsim 10^{-7}$ even if the
initial lepton asymmetries are taken to be smaller.  The only
difference is the temperature at which the asymmetries equilibrate
when $g =0$.

\begin{figure}
    \centering
    \includegraphics[width=10cm]{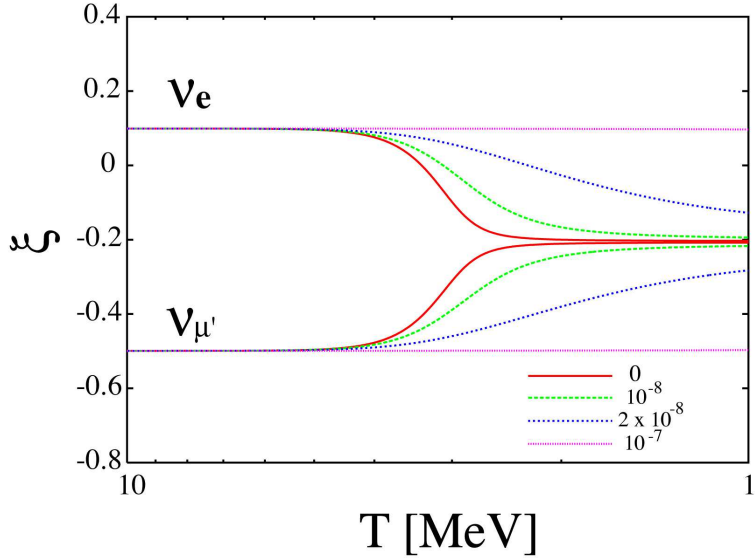}
    \caption{
The evolutions of $\xi_e$ and $\xi_{\mu'}$ for several values of $|g|$
with $\sin^2 \theta = 0.315$ and $\delta m_{21}^2 = 7.3 \times 10^{-5} {\rm eV}^2$. 
The initial conditions are $\xi_e = 0.1$ and $\xi_{\mu'}=-0.5$.
  }
       \label{fig:diagram7}
\end{figure}

We can repeat the same arguments for oscillations between $\nu_e$ and
$\nu_{\mu'}$.  The squared mass difference and mixing angle parameters
for the oscillations between $\nu_e$ and $\nu_{\mu'}$ are $\delta
m_{21}^2\equiv m_2^2-m_1^2= 7.3 \times 10^{-5}{\rm eV}^2$ and $\sin^2
\theta=0.315$~\cite{Fogli:2003kp}.  The coupling constant matrix
$g_{ab}$ is similarly approximated to be $g_{ab} = g \,\delta_{a \mu'}
\delta_{b \mu'}$.  Also the effective potential induced by the energy
densities of electrons and positrons is taken into account, since here
we consider oscillations including $\nu_{e}$.  The lower bound of
$|g|$ becomes slightly relaxed due to the smaller mass difference, as
can be seen from Fig.~\ref{fig:diagram7}.  The reason why the lepton
asymmetries of $\nu_e$ and $\nu_{\mu'}$ are not equilibrated
completely when $g=0$ is that the mixing angle we adopt is not
maximal.

Lastly, we comment on the form of the coupling matrix, $g_{ab}$, for
which the $(\nue-\nu_{\mu'})$-transformations are suppressed and,
consequently, a large lepton asymmetry of the universe is allowed.  So
far we have assumed that $g_{ab}$ is diagonal in order to deal with
both $\nu_\tau-\nu_\mu$ and $\nu_e-\nu_{\mu'}$ oscillations similarly.
However, in fact, a coupling matrix with more generic form works as
well.  We would like to stress that the equilibration of the muonic
and tauonic lepton asymmetries is not harmful to our purpose, as long
as these asymmetries are not erased through either oscillations or by
the reactions shown in Figs.~\ref{fig:diagram3} and
\ref{fig:diagram4}.  Therefore we do not have to assume the
hierarchical structure in $g_{\tau \tau}$, $g_{\mu \tau}$, and $g_{\mu
\mu}$, and they are constrained only by the astrophysical bounds:
\be
|g_{ab}| < 5 \times 10^{-6}
\ee
with $(a, b) = \{(\tau,\tau),(\mu, \tau), (\mu,\mu)\}$. In order to
suppress the oscillations between $\nu_e$ and $\nu_{\mu'}$, at least
one of these coupling constants should be larger than $\sim10^{-7}$,
while $|g_{ee}|$, $|g_{e\mu}|$, and $|g_{e\tau}|$ must be much smaller
than $\sim10^{-7}$. Thus we conclude that the coupling matrix suitable
for our purpose should satisfy:
\be
&|g_{ee}|,~|g_{e\mu}|,~|g_{e\tau}|& \ll 10^{-7},\non\\
10^{-7} \lesssim &{\rm Max}\,\left[\,|g_{\tau \tau}|,~|g_{\mu \tau}|,~|g_{\mu \mu}|\,\right]
 &\lesssim 5 \times 10^{-6}.
 \label{eq:form}
\ee
In the simplest class of majoron models, the coupling matrix,
$g_{ab}$, is proportional to the neutrino mass matrix $(m_\nu)_{ab}$.
In particular, for the normal mass hierarchy, $m_1 < m_2 < m_3$, the
reconstructed mass matrix is often parametrized
as~\cite{Smirnov:2003xe}
\be
g \propto m_\nu \propto
\left(
\begin{tabular}{lll}
0 & 0 & $\lambda$ \\
0 & 1 & 1\\
$\lambda$ & 1 & 1
\end{tabular}
\right)~ {\rm or}~
\left(
\begin{tabular}{lll}
$\lambda^2$ & $\lambda$ & $\lambda$\\
$\lambda$ & 1 & 1\\
$\lambda$ & 1 & 1\\
\end{tabular}
\right)
\ee
with $\lambda \sim 0.2$. Therefore the coupling matrix of this type
can satisfy the conditions (\ref{eq:form}). Of course, in the more
involved models, the coupling matrix is not necessarily proportional
to the neutrino mass matrix.


\section{Discussions and Conclusions \label{s-discuss}}

In the preceding sections we have shown that neutrino oscillations in
primordial plasma can be blocked by an introduction of the
neutrino-majoron interaction with a moderate value of coupling
constants, $|g| \lesssim 5 \cdot 10^{-6}$. Here let us take up the
case of $ |g| \gtrsim 10^{-5}$.  For a definite discussion, we assume
that the elements of coupling constant matrix whose indices involve
electronic neutrino are much smaller than the other elements,
i.e. $|g_{ee}|,~|g_{ea}| \ll |g_{ab}|$ with $(a, b) =
\{(\tau,\tau),(\mu, \tau), (\mu,\mu)\}$, so the electron type lepton
number conserves during the relevant BBN epoch. As we have seen in
section~\ref{s-react}, the muonic and/or tauonic lepton number is not
conserved for $|g_{ab}| \gtrsim 10^{-5}$. The fate of the large
degeneracy in $\nu_\mu$ and $\nu_\tau$ depends on whether this
$L$-violating interaction reached equilibrium before or after
$\nu_\mu$ and $\nu_\tau$ decoupled from the thermal bath.

First let us consider the former scenario. Since $\nu_\mu$ and
$\nu_\tau$ kept on to be in strong contact with the primeval plasma
when the $L$-violating interactions were efficient, the muonic and/or
tauonic lepton asymmetry vanished and thereby heated the plasma.  At
the same time majorons were abundantly produced and contributed to the
effective number of neutrino species as $\Delta N_\nu = 4/7$. The CP
asymmetric part of the effective potential of neutrino due to the
muonic and tauonic lepton asymmetries vanished, but there would be a
similar (but possibly smaller) term due to the charge asymmetry of
$\nu_e$ (see Eqs.~(\ref{etanue}) and (\ref{etanumu})). Also there
exists the contribution the the effective potential due to abundant
majorons in the plasma (see Eq.~(\ref{eq:eff_potential2})).  A further
important point is that $L$-violating processes, shown in
Fig.~\ref{fig:diagram3}, induces the effective potential between
neutrino and antineutrino, since $\langle a^\dagger b \rangle$ no
longer vanishes. However, neutrino oscillations between $\nu_{e}$ and
$\nu_{\mu'}$ can be treated in the same way as before, since electron
type lepton number remains a well conserved quantity.  Therefore, we
can deduce from the previous results that neutrino oscillations
between $\nu_{e}$ and $\nu_{\mu'}$ are blocked and $\xi_e$ remains
intact.  It should be noted that equilibrium majorons, instead of
excessive $\nu_\mu$ and $\nu_\tau$, contribute to the extra radiation,
which speeds up the expansion rate, and that its contribution, $\Delta
N_\nu = 4/7$, well satisfies the constraint shown in
Eq.~(\ref{extra_rad}).

Next we discuss the other possibility that the $L$-violating
interactions come to thermal equilibrium after decoupling of $\nu_\mu$
and $\nu_\tau$.  Majorons are also produced, as in the previous case.
However, since $\nu_\mu$ and $\nu_\tau$ cannot exchange energy with
the primeval plasma, the total energy of majorons, $\nu_\mu$ and
$\nu_\tau$ must be conserved. Therefore the excessive energy
previously stored in $\nu_\mu$ and $\nu_\tau$ is redistributed among
majorons, $\nu_\mu$, and $\nu_\tau$, and their distribution would be
like Eq.~(\ref{fnu=fbar}) but without any charge asymmetry, i.e. with
$\xi = \bar\xi$.  It is thus clear that $\Delta N_\nu$ remains
unchanged before and after $L$-violating interaction comes into
equilibrium.  Our concern here is whether the modified distribution of
$\nu_{\mu'}$ affects that of $\nu_e$ through neutrino oscillations.
However, the oscillations are blocked on the same ground.

In connection with the second case, there is an interesting
possibility that the majoron-exchange reactions: $\nu_{\mu, \tau}
\bar\nu_{\mu,\tau} \leftrightarrow \nu_e \bar\nu_e$, $\nu_{\mu,
\tau}\nu_{\mu, \tau} \leftrightarrow \nu_e \nu_e$, $\nu_{\mu,
\tau}\nu_{\mu, \tau} \leftrightarrow \bar\nu_e \bar\nu_e$,
$\bar\nu_{\mu, \tau}\bar\nu_{\mu, \tau} \leftrightarrow \nu_e \nu_e$
and $\bar\nu_{\mu, \tau}\bar\nu_{\mu, \tau} \leftrightarrow \bar\nu_e
\bar\nu_e$, could produce additional electronic neutrinos and
antineutrinos. These processes can proceed if $|g_{ee}|$ and
$|g_{ea}|$ with $a= \mu,~\tau$ are sizable $\gtrsim 10^{-5}$.  If this
is the case, the relative abundance of $\nu_e$ and $\bar\nu_e$ with
respect to photons and electrons would be larger than in the standard
model. This would lead to a later neutron-proton freezing and to a
smaller number density of survived neutrons, which might compensate
the effect of additional energy density stored in all species of
neutrinos and majorons and even overshoot it.  Then we do not need to
suppress neutrino oscillations, therefore the coupling constant matrix
$g_{ab}$ is allowed to take rather arbitrary values as long as the
$L$-violating processes come into equilibrium after decoupling of
muonic and tauonic neutrinos.

Thus our examination indicates that a conspiracy between the speed-up
effect and the shift of $\beta$ equilibrium is possible for $ |g_{ab}|
\gtrsim 10^{-5}$, as long as $|g_{ee}|,~|g_{ea}| \ll |g_{ab}|$ with
$a,~b = \mu,~\tau$. Furthermore, even for $|g_{ee}|,~|g_{ea}| \gtrsim
10^{-5}$, such a cancellation might be possible in a somewhat
different way.  This considerably extends the allowed parameter space,
although rigorous study might be necessary to obtain further
quantitative results.

In this paper we have shown that the hypothetical neutrino-majoron
interaction can suppress neutrino oscillations in the primordial
plasma to prevent lepton asymmetries of all neutrino species from
being equilibrated. The exact form of the effective potential induced
by this interaction is calculated. We have found an allowed range of
the coupling constant: $10^{-7} < |g| < 5\cdot 10^{-6}$ (more
precisely, see Eq.~(\ref{eq:form})), which satisfies the astrophysical
bounds and makes the scenario operative.  For the coupling constant in
this range, $\nu_e-\nu_{\mu'}$ oscillation in the early Universe is
blocked, thereby keeping the cosmological lepton asymmetry of electron
type unchanged.  The upper bound comes from the requirement that
lepton number is effectively conserved, and the lower bound is
obtained from the study of the evolution of the lepton asymmetries
both analytically and numerically, in two flavor approximation.  The
constant matrix in the simplest class of majoron models can satisfy
the desired constraints, in the case of the normal mass hierarchy.
Furthermore, even for $ |g| \gtrsim 10^{-5}$, the large energy density
necessary to cancel the effect of $\xi_e$ can be supplied by majorons,
$\nu_\mu$ and $\nu_\tau$, even if muonic and/or tauonic lepton
asymmetries were erased. Thus we conclude that an addition of the
majoron field to the standard model can reopen a possibility that the
effect of $\xi_e$ is compensated by large $\xi_{\mu,\tau}$ (or by the
extra energy of majoron itself), thereby curing a probable discrepancy
between the BBN and CMBR.

\appendix
\section{Derivation of effective potential due to majoron-neutrino interaction}
\label{app:der}
Here we present the derivation of the effective potentials Eqs.~(\ref{eq:eff_potential})
and (\ref{eq:eff_potential2}).
First the notations and conventions 
that we adopt here, are listed. The metric is taken
to be $(-,+,+,+)$. The gamma matrices in the chiral representation are
\begin{eqnarray}
\label{eq:gam_mat_chiral}
\gamma^0 = -i \left(
\begin{array}{cc}
0&1\\
1&0
\end{array}
\right),
~~~
\gamma^i = -i \left(
\begin{array}{cc}
0&\sigma_i\\
-\sigma_i&0
\end{array}
\right),
~~~
\gamma_5= \left(
\begin{array}{cc}
1&0\\
0&-1
\end{array}
\right),
\end{eqnarray}
where $\sigma_i$ are the Pauli matrices. The gamma matrices in the Dirac representation
are related to those in the chiral representation as
\begin{eqnarray}
\gamma^\mu_{\rm Dirac} = U \gamma^\mu_{\rm chiral} U^\dagger,
~~~
U= \frac{1}{\sqrt{2}}\left(
\begin{array}{cc}
1&1\\
1&-1
\end{array}
\right),
\end{eqnarray}
and they are given by 
\begin{eqnarray}
\label{eq:gam_mat_Dirac}
\gamma^0 = -i \left(
\begin{array}{cc}
1&0\\
0&-1
\end{array}
\right),
~~~
\gamma^i = i \left(
\begin{array}{cc}
0&\sigma_i\\
-\sigma_i&0
\end{array}
\right),
~~~
\gamma_5= \left(
\begin{array}{cc}
0&1\\
1&0
\end{array}
\right).
\end{eqnarray}
The charge conjugation matrix $C$ is defined as
\be
C \equiv i \gamma^2 \gamma^0 = \left\{
\begin{array}{ll}
-i
\left(
\begin{array}{cc}
\sigma_2&0\\
0&-\sigma_2
\end{array}
\right)
& {\rm (chiral)}\\
&\\
-i
\left(
\begin{array}{cc}
0&\sigma_2\\
\sigma_2&0
\end{array}
\right)
& {\rm (Dirac)} 
\end{array}
\right..
\ee
Hereafter we work with the gamma matrices in the Dirac representation.
The momentum expansion of a  free Dirac field $\psi(x)$ with mass $m$ is given by
\begin{eqnarray}
\psi(x)&=& \sum_s \int d {\bf p} \left(
a({\bf p},s) u({\bf p},s) e^{-i E_{\bf p} t + i {\bf p}\,{\bf x}}
+ b^\dagger({\bf p},s) v({\bf p},s)e^{i E_{\bf p} t - i {\bf p}\,{\bf x}}\right),\non\\
 &=& \sum_s \int d {\bf p} \left(
a({\bf p},s) u({\bf p},s) + b^\dagger(-{\bf p},s) v(-{\bf p},s)\right)
e^{-i p^0 t + i {\bf p}\,{\bf x}},
\end{eqnarray}
where we defined 
$p^0 \equiv \epsilon_p E_{\bf p}= \epsilon_p\sqrt{{\bf p}^2 + m^2}$
with $\epsilon_p = \pm 1$. The sign of $\epsilon_p$ is chosen
so that  the positive and negative-energy solutions are reproduced.
$s=\pm$ describes the sign of the spin projection eigenvalue,
and we have used the notation
$d {\bf p} \equiv d^3 {\bf p}/(2 \pi)^3$.
The Dirac spinors $u({\bf p},s)$ and $v({\bf p},s)$ take the
form,
\begin{eqnarray}
u({\bf p},s) &=& \frac{1}{\sqrt{2 E_{\bf p} (E_{\bf p}+m)}}
\left(
\begin{array}{c}
E_{\bf p}+m\\
-{\bf p} \cdot {\bf \sigma}
\end{array}
\right) \chi_u(s),\non\\
v({\bf p},s) &=& \frac{1}{\sqrt{2 E_{\bf p} (E_{\bf p}+m)}}
\left(
\begin{array}{c}
-{\bf p} \cdot {\bf \sigma}\\
E_{\bf p}+m
\end{array}
\right) \chi_v(s).
\end{eqnarray}
The two-component spinors $\chi_u(s)$ and $\chi_v(s)$ are not independent
but related to each other 
through: $\chi_v(s) = - i \sigma_2 \chi_u(s)^*$. If we choose
them to be eigenstates of the helicity operator ${\bf p \cdot \sigma}/|{\bf p}|$, they are
\begin{eqnarray}
\chi_u(+) &=& \left(
\begin{array}{c}
e^{-i \varphi/2} \cos{ \frac{\theta}{2} }\\
e^{i \varphi/2} \sin{ \frac{\theta}{2} }
\end{array}
\right),~~~
\chi_u(-) = \left(
\begin{array}{c}
-e^{-i \varphi/2} \sin{ \frac{\theta}{2} }\\
e^{i \varphi/2} \cos{ \frac{\theta}{2} }
\end{array}
\right),\non\\
\chi_v(+) &=& \left(
\begin{array}{c}
-e^{-i \varphi/2} \sin{ \frac{\theta}{2} }\\
e^{i \varphi/2} \cos{ \frac{\theta}{2} }
\end{array}
\right),~~~
\chi_v(-) = \left(
\begin{array}{c}
-e^{-i \varphi/2} \cos{ \frac{\theta}{2} }\\
-e^{i \varphi/2} \sin{ \frac{\theta}{2} }
\end{array}
\right),
\end{eqnarray}
where $\theta$ and $\varphi$ are defined as
${\bf p} = |{\bf p}|(\sin\theta \cos\varphi,\sin\theta \sin\varphi,\cos\theta)$. 
For flipped momentum $-{\bf p}$, $\{\theta, \varphi\}$ should be 
replaced with $\{\pi-\theta, \varphi+\pi\}$.

In deriving the effective potential, the neutrino field $\nu(x)$ is approximated
to be a massless left-handed
field :
\begin{eqnarray}
\nu_i(x) &=& \int d {\bf p} \left(
a_i({\bf p}) u_{\bf p} + b_i({-{\bf p}})^\dagger v_{-{\bf p}}
\right)
e^{-i p^0 t + i {\bf p}\,{\bf x}},
\end{eqnarray}
where
\begin{eqnarray}
u_{\bf p} &\equiv& \frac{1+\gamma_5}{2} u({\bf p},-) = \frac{1}{\sqrt{2}} 
\left(
\begin{array}{c}
1\\
1
\end{array}
\right) \chi_u(-),\non\\
u_{-{\bf p}} &\equiv& \frac{1+\gamma_5}{2} u(-{\bf p},-) = \frac{i}{\sqrt{2}} 
\left(
\begin{array}{c}
1\\
1
\end{array}
\right) \chi_u(+),\non\\
v_{{\bf p}} &\equiv&  \frac{1+\gamma_5}{2} v({\bf p},+)=
\frac{1}{\sqrt{2}} 
\left(
\begin{array}{c}
1\\
1
\end{array}
\right) \chi_u(-),\non\\
v_{-{\bf p}} &\equiv&  \frac{1+\gamma_5}{2} v(-{\bf p},+)=
\frac{i}{\sqrt{2}} 
\left(
\begin{array}{c}
1\\
1
\end{array}
\right) \chi_u(+).
\end{eqnarray}
The annihilation and creation operators $a$, $b$, $a^\dagger$ and $b^\dagger$ satisfy 
the anticommutation relations.
\be
\label{eq:anticom}
\left\{a_i({\bf p}),\,a_j({\bf q})^\dagger \right\} &=& (2 \pi)^3 \delta_{ij}\, \delta^{(3)}({\bf p}-{\bf q}),\non\\
\left\{b_i({\bf p}),\,b_j({\bf q})^\dagger \right\} &=& (2 \pi)^3 \delta_{ij}\, \delta^{(3)}({\bf p}-{\bf q}),\non\\
{\rm others} &=& 0.
\ee
Similarly, the momentum expansion of the free majoron field is 
\begin{eqnarray}
\chi(x) &=& \int d {\bf p} \left(
a_\chi({\bf p}) u_\chi({\bf p}) + a_\chi({-{\bf p}})^\dagger v_\chi({-{\bf p}})
\right)
e^{-i p^0 t + i {\bf p}\,{\bf x}},
\ee
where
\be
u_\chi({\bf p}) =v_\chi({\bf p}) = \frac{1}{\sqrt{2 E_{\bf p}}}.
\end{eqnarray}
The annihilation and creation operators $a_\chi$, $a_\chi^\dagger$ satisfy 
the commutation relations:
\be
\label{eq:anticom2}
\left[a_\chi({\bf p}),\,a_\chi({\bf q})^\dagger \right] &=& (2 \pi)^3 \, \delta^{(3)}({\bf p}-{\bf q}),\non\\
{\rm others} &=& 0.
\ee

Next we derive the effective Hamiltonian Eq.~(\ref{eq:eff_hamiltonian}). The equation of motion for the
majoron field $\chi$ is 
\be
\label{eq:eom_for_majoron}
\del_\mu \del^\mu \chi =- \frac{i}{2}  \left( g_{ab}\, \nu_a ^T C \nu_b 
                      					  +g_{ab}^* \, \nu_b^\dagger C \nu_a^*
 							\right).
\ee
To solve this equation, we define the majoron and neutrino fields in momentum space :
\be
\chi(x) &=& \int \frac{d^4 p}{(2 \pi)^4}\, \tilde{\chi}(p)\, e^{i p x},\non\\
\nu_a(x) &=& \int \frac{d^4 p}{(2 \pi)^4}\, \tilde{\nu}_a(p)\, e^{i p x}.
\ee
Then the solution of Eq.~(\ref{eq:eom_for_majoron}) is
\be
\label{eq:sol}
\tilde{\chi}(p) &=& \frac{i}{2 p^2} \int  \frac{d^4 q}{(2 \pi)^4}
\left\{
 g_{ab}\, \tilde{\nu}_a(p-q)^T C \tilde{\nu}_b(q) 
        +g_{ab}^* \, \tilde{\nu}_b (-p+q)^\dagger C\tilde{ \nu}_a^*(-q)
\right\}.
\ee
Thus the effective Hamiltonian is:
\be
{\cal H}_{\rm eff}^{(\nu\nu)} &=&
-\int d^3 {\bf x} \, {\cal L}_{\rm eff}^{(\nu\nu)},\non\\
&=&  -\frac{1}{2}\int d^3 {\bf x} \,\frac{i}{2} 
\int \frac{d^4 p}{(2 \pi)^4}\, e^{i p x} 
 \frac{i}{2 p^2} \int  \frac{d^4 q}{(2 \pi)^4}
\left\{
 g_{ab}\, \tilde{\nu}_a(p-q)^T C \tilde{\nu}_b(q) \right.\non\\
 &&\left.
        +g_{ab}^* \, \tilde{\nu}_b (-p+q)^\dagger C\tilde{ \nu}_a^*(-q)
\right\}  \int \frac{d^4 r}{(2 \pi)^4}\, e^{i r x} \int \frac{d^4 s}{(2 \pi)^4}\, e^{i s x} \non\\
&&
\times  \left\{
 g_{cd}\, \tilde{\nu}_c(r)^T C \tilde{\nu}_d(s) 
        +g_{cd}^* \, \tilde{\nu}_d (-s)^\dagger C\tilde{ \nu}_c^*(-r)
        \right\}.
\ee
where we substituted Eq.~(\ref{eq:sol}) into the second equation. 
Note that the numerical factor $1/2$
is inserted  in front of the last equation.
In a first-order perturbative
approximation the neutrino and majoron fields can be set to be free fields:
\be
\tilde{\nu}_i (p) &=& 2 \pi \delta (p^0 - \epsilon_p E_{\bf p}) \,\nu_i({\bf p}),\non\\
\tilde{\chi}(p) &=& 2 \pi \delta (p^0 - \epsilon_p E_{\bf p}) \,\chi({\bf p}),
\ee
where
\be
\nu_i({\bf p}) &=& a_{i}({\bf p}) u_{\bf p} + b_{i}(-{\bf p})^\dagger v_{-{\bf p}},\non\\
\chi({\bf p}) &=& a_{\chi}({\bf p}) u_\chi({\bf p}) + a_\chi(-{\bf p})^\dagger v_\chi({-{\bf p}}).
\ee
Thus the effective Hamiltonian becomes
\be
{\cal H}_{\rm eff}^{(\nu\nu)} (t) &=&
 -\frac{1}{16} \int d{\bf p} \,d{\bf q}\, d{\bf r}\, d{\bf s}\,\, (2 \pi)^3 
\delta^{(3)}({\bf p+q+r+s}) 
\,e^{-i E_{\rm tot} t} \non\\
&&~~~~~~~~
 \times \frac{F({\bf p}, {\bf q})F({\bf r}, {\bf s})}{\epsilon_p \epsilon_q E_{\bf p} E_{\bf q} (1-
\epsilon_p \epsilon_q
\cos \theta_{\bf pq})},
\ee
where 
\be
E_{\rm tot} &=& \epsilon_p E_{\bf p} +  \epsilon_q E_{\bf q} + \epsilon_r E_{\bf r} + \epsilon_s E_{\bf s} ,\non\\
\cos \theta_{\bf pq} & \equiv & \frac{{\bf p} \cdot {\bf q}}{\left|{\bf p} \right| \left|{\bf q} \right|},\non\\
F({\bf p}, {\bf q})& \equiv &  g_{ab}\, \nu_a({\bf p}) ^T C \nu_b({\bf q}) 
+g_{ab}^* \, \nu_b({\bf -q})^\dagger C \nu_a^*({\bf -p}).
\ee

Once the effective Hamiltonian is obtained, we can calculate the effective potential for neutrinos
according to Ref.~\cite{Sigl:fn}.  All we have to do is to evaluate
\be
\la\left[
{\cal H}_{\rm eff} (t=0), a_j({\bf p})^\dagger a_i({\bf p})
\right]\ra &=& - (2 \pi)^3 \delta^{(3)}(0)\left[V_{\bf p},  \rho_{\bf p}\right]_{i j},
\label{eq:sigl_raffelt}
\ee
where $\left[V_{\bf p} \right]_{ab}$ is the effective potential matrix. Using the anticommutation relations~(\ref{eq:anticom}), we obtain the final result:
\be
\left[V_{\bf p}^{(\nu\nu)}\right]_{ab} = \int d{\bf q}~~\frac{1}{4 \left|{\bf p} \right| \left|{\bf q} \right|} \left[
g^\dagger \left(\rho^T_{\bf q} + \overline{\rho}^T_{\bf q} \right) g
\right]_{ab}.
\ee

Next we derive the effective Hamiltonian, which describes neutrino-majoron 
scattering. In this case we integrate out the neutrino field instead of majoron field.
The equation of motion for the neutrino is
\be
-i \gamma^0 \gamma^\mu \partial_\mu \nu_a + i \chi g_{ab}^{*} C \nu_b^*=0.
\ee
The solution is then
\be
\tilde{\nu}_a(p)=  \frac{i p_\mu}{p^2}
g_{ab}^* \int \frac{d^4 q}{(2 \pi)^4} \tilde{\chi}(p+q) \gamma^\mu 
\gamma^0 C \tilde{\nu}_b^*(q).
\ee
Substituting this result into Eq.~(\ref{eq:maj-nu-int}), we obtain the effective Hamiltonian:
\be
{\cal H}_{\rm eff}^{(\nu\chi)}(t) &=&
-\frac{i}{2}\, g_{ab}g_{bc}^* \int d^3 {\bf x}
\int  \frac{d^4 p}{(2 \pi)^4}   \frac{d^4 q}{(2 \pi)^4}
  \frac{d^4 r}{(2 \pi)^4}   \frac{d^4 s}{(2 \pi)^4}
\,e^{i (p+q+r+s) x} \non\\
&& \times\, \tilde{\chi}(p)\tilde{\chi}(q)\, \overline{\tilde{\nu}}_c(-r) \gamma^\mu
\tilde{\nu}_a(s)\, \frac{(q+r)_\mu}{(q+r)^2},\non\\
&=& -\frac{i}{2}\, \left[g^\dagger g\right]_{ab}
 \int d{\bf p} \,d{\bf q}\, d{\bf r}\, d{\bf s}\,\, (2 \pi)^3 
\delta^{(3)}({\bf p+q+r+s}) 
\,e^{-i E_{\rm tot} t} \non\\
&&\times \chi({\bf p})\chi({\bf q})
\overline{\nu}_a(-{\bf r}) \gamma^\mu
\nu_b({\bf s})\, \left.\frac{(q+r)_\mu}{2 q \cdot r}\right|_{
q^0=\epsilon_q E_{\bf q},r^0=\epsilon_r E_{\bf r}
},
\ee
where we take both neutrino and majoron fields to be on-shell in the second equation.
Applying Eq.~(\ref{eq:sigl_raffelt}), we obtain the effective potential 
due to the neutrino-majoron
scattering:
\be
\left[V_{\bf p}^{(\nu\chi)}\right]_{ab} =\int d{\bf q}~~\frac{f_{\chi}({\bf q})}{
4 \left|{\bf p} \right| \left|{\bf q} \right|} \left[ g^\dagger g\right]_{ab},
\ee
where the number density of majorons 
with momentum ${\bf p}$, $n_{\bf p}^{(\chi)}$,
is defined as
\be
\la a^\dagger_\chi({\bf p}) \,a_\chi({\bf p}')\ra &=& (2 \pi)^3 \delta^{(3)}({\bf p}-{\bf p}') \,
f_{\chi}({\bf p}).
\ee

\section{Invariant amplitude squared for $2\nu \leftrightarrow 2 \bar\nu$}
\label{app:amplitude}

Here we calculate the invariant amplitude squared for the reaction, $2\nu \leftrightarrow 2 \bar\nu$.
Provided that  the diagonal coupling constants are larger than 
the non-diagonal ones,
and that one of the diagonal coupling constants dominates, 
we consider neutrinos of the
flavor with the largest coupling. Hereafter we drop the sub-indices for flavor.
The diagrams, which contribute to the reaction, are shown in Fig.~\ref{fig:diagram3}.

The S-matrix and the invariant amplitude squared are related as 
\be
S_{\beta \alpha}&=& 2 \pi (-i) \delta^{(4)}(p_\beta - p_\alpha) {\cal M}_{\beta \alpha},\non\\
\left|M_{\rm inv}\right|^2 &=& (2 \pi)^{-6} \prod_{\alpha}\left[(2\pi)^3 2E_\alpha \right]
        					    \prod_{\beta}\left[(2\pi)^3 2 E_\beta\right]
					     \left| {\cal M}_{\beta \alpha}\right|^2,
\label{eq:def_for_M}
\ee
where $\alpha$ and $\beta$ represent symbolically the initial and 
final states . 
Each matrix element for s-, t- and u-channel diagrams is given as
\be
{\cal M}^{(s)}_{p_3 p_4,p_1  p_2} &=& \frac{g^2}{(2 \pi)^3} \frac{1}{(p_1 + p_2)^2}
       			(u_{{\bf p}_2}^T C u_{{\bf p}_1}) (v_{{\bf p}_4}^T C v_{{\bf p}_3}),\non\\
{\cal M}^{(t)}_{p_3 p_4,p_1  p_2} &=& -\frac{g^2}{(2 \pi)^3} \frac{1}{(p_1 - p_3)^2}
       			(v_{{\bf p}_3}^T C u_{{\bf p}_1}) (v_{{\bf p}_4}^T C u_{{\bf p}_2}),\non\\	
{\cal M}^{(u)}_{p_3 p_4,p_1  p_2} &=& \frac{g^2}{(2 \pi)^3} \frac{1}{(p_1 - p_4)^2}
       			(v_{{\bf p}_4}^T C u_{{\bf p}_1}) (v_{{\bf p}_3}^T C u_{{\bf p}_2}),
\ee
where the four momenta satisfy the conservation condition, 
$p_1+p_2 = p_3+p_4$.  If we use the explicit
expressions for $u_{\bf p}$ and $v_{\bf p}$ derived in the previous section,
we obtain
\be
{\cal M}^{(s)}_{p_3 p_4,p_1  p_2} +
{\cal M}^{(t)}_{p_3 p_4,p_1  p_2} +
{\cal M}^{(u)}_{p_3 p_4,p_1  p_2}  = \frac{g^2}{(2 \pi)^3} \frac{1}{\sqrt{E_1 E_2 E_3 E_4}}\,
\frac{3}{4}.
\ee
Substituting this result into Eq.~(\ref{eq:def_for_M}), the invariant amplitude squared 
becomes 
\be
\left|M_{\rm inv}\right|^2 &=& 9 |g|^4.
\ee

\section{Formulas for the collisional integral in the Boltzmann approximation}
\label{app:formula}
Here we write down the formulas for the collisional 
integral in the Boltzmann approximation for completeness.
For the derivation, see Ref.~\cite{Dolgov:1997mb}.
\be
\int \frac{d^3{\bf y}_2}{y_2}\int \frac{d^3{\bf y}_3}{y_3}\int \frac{d^3{\bf y}_4}{y_4} e^{-y_3-y_4} \delta^{(4)}(y_1+y_2-y_3-y_4)
&=& 4 \pi^2 e^{-y_1},\non\\
\int \frac{d^3{\bf y}_1}{y_1}\int \frac{d^3{\bf y}_2}{y_2}\int \frac{d^3{\bf y}_3}{y_3}
\int \frac{d^3{\bf y}_4}{y_4} e^{-y_3-y_4} \delta^{(4)}(y_1+y_2-y_3-y_4)
&=& 32 \pi^3.
\ee

\bigskip

{\bf Acknowledgment}

A.D. is grateful to the Research Center for the Early Universe of the
University of Tokyo for the hospitality during the time when this work
started.  F. T. is grateful to M. Kawasaki and K. Ichikawa for useful
discussions.  F. T. would like to thank the Japan Society for the
Promotion of Science for financial support.


\begin{thebibliography}{99}

\bibitem{dolgov02}
A.~D.~Dolgov,
Phys.\ Rept.\  370 (2002) 333.

\bibitem{kohri96}
K.~Kohri, M.~Kawasaki and K.~Sato,
Astrophys.\ J.\  490 (1997) 72.

\bibitem{Barger:2003zg}
V.~Barger, J.~P.~Kneller, H.~S.~Lee, D.~Marfatia and G.~Steigman,
Phys.\ Lett.\ B 566 (2003) 8.

\bibitem{hansen01}
S.H.~Hansen, G.~Mangano, A.~Melchiorri, G.~Miele and O.~Pisanti,
Phys.\ Rev.\ D 65 (2002) 023511.



\bibitem{largeL_smallB}
J.~A.~Harvey and E.~W.~Kolb, Phys.\ Rev.\ D 24 (1981) 2090;\\
A.D. Dolgov, D.K. Kirilova, J. Moscow Phys. Soc. 1, (1991), 217;\\
A.D. Dolgov, Phys. Repts, 222 (1992) No. 6;\\
A.~Casas, W.~Y.~Cheng and G.~Gelmini,
Nucl.\ Phys.\ B 538 (1999) 297;\\
J.~March-Russell, H.~Murayama and A.~Riotto,
JHEP 9911 (1999) 015;\\
J.~McDonald,
Phys.\ Rev.\ Lett.\  84 (2000) 4798;\\
M.~Kawasaki, F.~Takahashi and M.~Yamaguchi,
Phys.\ Rev.\ D 66 (2002) 043516;\\
M.~Yamaguchi,
Phys.\ Rev.\ D 68 (2003) 063507;\\
T.~Chiba, F.~Takahashi and M.~Yamaguchi,
Phys.\ Rev.\ Lett.\  92 (2004) 011301;\\
F.~Takahashi and M.~Yamaguchi,
arXiv:hep-ph/0308173, Phys.\ Rev.\ D (to be published).

\bi{nu-mixing}
SNO Collaboration, Phys. Rev. Lett. 89 (2002) 011301; 
Phys. Rev. Lett. 89 (2002) 011302;\\
KamLAND Collaboration, Phys. Rev. Lett. 90 (2003) 021802;\\
S. Fukuda {\it et al.}, Super-Kamiokande Coll.,
Phys.\ Lett.\ B{539} (2002) 179;\\
B.T. Cleveland {\it et al.}, Astrophys. J. {496}, 505 (1998);\\
D.N. Abdurashitov {\it et al.}, SAGE Coll.,
Phys. Rev. {C60}, 055801 (1999); astro-ph/0204245;\\
W. Hampel {\it et al.}, GALLEX Coll., Phys. Lett. {B447}, 127 (1999);\\
C. Cattadori, GNO Coll., Nucl.\ Phys.\ B (Proc.\ Suppl.) {110} (2002) 311.\\
Super-Kamiokande Coll., Y. Fukuda {\it et al.}, 
Phys. Rev. Lett. {81} (1998) 1562;\\
MACRO Coll., M. Ambrosio {\it et al.}, Phys. Lett. B{434} (1998) 451.

\bi{act-act}
A.D.~Dolgov, S.H.~Hansen, S.~Pastor, S.T.~Petcov, G.G.~Raffelt,
D.V.~Semikoz,  Nucl.Phys. B632 (2002) 363.

\bi{lunardini01}
C.Lunardini, A.Yu.Smirnov, Phys. Rev. D64 (2001) 073006;\\
Y.Y.Y. Wong, Phys.Rev. D66 (2002) 025015;\\
K.N. Abazajian, J.F. Beacom, N.F. Bell, Phys.Rev. D66 (2002) 013008.

\bibitem{Barger:2003rt}
V.~Barger, J.~P.~Kneller, P.~Langacker, D.~Marfatia and G.~Steigman,
Phys.\ Lett.\ B 569 (2003) 123.

\bi{majoron}
Y. Chikashige, R.N.~Mohapatra, R.D.~Peccei,
Phys. Rev. Lett.  {45} (1980) 1926;\\
G.B.~Gelmini, M.~Roncadelli,
Phys. Lett. B{99} (1981) 411;\\
H.M. Georgi, S.L. Glashow, S. Nussinov, Nucl. Phys. B193 (1981) 297;\\
A.Yu. Smirnov, Yad. Fiz. 34 (1981) 1547;\\
J. Schechter, J.W.F. Valle,
Phys. Rev. D{25} (1982) 774.

\bibitem{babu92}
K.S. Babu and I.Z. Rothstein, 
Phys. Lett. B{275} (1992) 112.

\bibitem{bento01}
L.~Bento and Z.~Berezhiani,
Phys. Rev. D{64} (2001) 115015.

\bi{ad-itep02}
A.D. Dolgov, Surveys High Energ. Phys. 17 (2002) 91.

\bi{nora}
D.~N\"otzold and G.~Raffelt, Nucl. Phys. B{307} (1988) 924.

\bi{v-off}
J. Pantaleone, Phys. Lett. B{287} (1992) 128;\\
S. Samuel, Phys.\ Rev.\ D{48} (1993) 1462;\\
V.A. Kosteleck{\'y}, J. Pantaleone, S. Samuel,
Phys.\ Lett.\ B{315} (1993) 46;\\
V.A. Kosteleck{\'y}, S. Samuel, Phys. Rev. D{49} (1994) 1740;\\
V.A. Kosteleck{\'y}, S. Samuel, Phys.\ Rev.\ D{52} (1995) 3184;\\
S. Samuel, Phys.\ Rev.\ D{53} (1996) 5382;\\
V.A. Kosteleck{\'y}, S. Samuel, { Phys. Lett.} { B385} (1996) 159;\\
S.~Pastor, G.G.~Raffelt, D.V.~Semikoz, Phys. Rev. D{65} (2002) 053011.



\bi{tomas01}
R. Tomas, H. P{\"a}s, J.W.F. Valle, Phys. Rev. D64 (2001) 095005.

\bi{farzan02}
Y. Farzan, Phys. Rev. D67 (2003) 073015.

\bi{raffelt}
G.G. Raffelt, {\it Stars as Laboratories for Fundamental Physics},
University of Chicago Press, 1996.

\bi{2-beta}
T. Bernatowicz {\it et al}, Phys. Rev. Lett. 69 (1992) 2341.


\bibitem{Beacom:2002vi}
J.~F.~Beacom, N.~F.~Bell, D.~Hooper, S.~Pakvasa and T.~J.~Weiler,
Phys.\ Rev.\ Lett.\  {90} (2003) 181301;\\
S.~Ando,
Phys.\ Lett.\ B 570 (2003) 11;\\
G.~L.~Fogli, E.~Lisi, A.~Mirizzi and D.~Montanino,
arXiv:hep-ph/0401227.

\bibitem{Ahrens:2003ix}
J.~Ahrens  [IceCube Collaboration],
arXiv:astro-ph/0305196.

\bibitem{bb02}
J.F. Beacom and N.F. Bell, Phys. Rev. D 65 (2002) 113009.

\bibitem{Sigl:fn}
G.~Sigl and G.~Raffelt, Nucl.\ Phys.\ B 406, 423 (1993).

\bi{adS}
A.D. Dolgov, S. H. Hansen, D. V. Semikoz, Nucl. Phys. B524 (1998) 621.

\bi{kin-eq}
A.D. Dolgov, Yad.Fiz.  33 (1981) 1309; English translation: Sov. J.
Nucl. Phys. 33 (1981) 700;\\
M.A. Rudzsky, Astrophys. Space Sci. 165 (1990) 65;\\
B.H.J.\ McKellar and M.J.\ Thomson, Phys.\ Rev.\ D 49 (1994) 2710.

\bi{ad-fv-03}
A.D. Dolgov, F.L. Villante,  Nucl. Phys. B679 (2004) 261.

\bibitem{Fogli:2003kp}
G.~L.~Fogli, E.~Lisi, A.~Marrone, D.~Montanino, A.~Palazzo and A.~M.~Rotunno,
eConf  C030626 (2003) THAT05
[arXiv:hep-ph/0310012].

\bibitem{Smirnov:2003xe}
A.~Y.~Smirnov,
arXiv:hep-ph/0311259.


\bibitem{Dolgov:1997mb}
A.~D.~Dolgov, S.~H.~Hansen and D.~V.~Semikoz,
Nucl.\ Phys.\ B  503 (1997) 426.

\end{thebibliography}
\end{document}